\documentclass[reprint,twocolumn,superscriptaddress,amsmath,amssymb,nofootinbib,preprintnumbers]{revtex4-2} 

\usepackage{mathptmx}
\usepackage{graphicx}  
\usepackage[mathlines]{lineno}
\usepackage[english]{babel} 
\usepackage{xcolor}
\usepackage{ulem}
\usepackage{dcolumn}   
\usepackage{bm}        
\usepackage{amssymb}   
\usepackage{amsbsy}   
\usepackage{comment}
\usepackage{multirow}
\usepackage{diagbox}
\usepackage[draft]{todonotes}
\usepackage{mathtools}
\usepackage{newtxtext}
\usepackage{upgreek}
\usepackage[varvw]{newtxmath}
\usepackage{nicefrac}
\usepackage{tikz-feynman}
\usepackage{feynmp-auto}
\usepackage{pgf}
\usepackage{tikz} 
\usetikzlibrary{shapes,arrows,automata}
\tikzfeynmanset{compat=1.1.0}

\makeatletter
\tikzfeynmanset{/tikzfeynman/every boson@@/.style={
    /tikz/draw=none,
    /tikz/decoration={name=none},
    /tikz/postaction={
      /tikz/draw,
      /tikz/decoration={
        zigzag,segment length=1mm, 
        amplitude=0.6mm,
      },
      /tikz/decorate=true,
    }
}}
\makeatother

\usepackage{hyperref}
\hypersetup{
    colorlinks=true,
    linkcolor=cyan,
    filecolor=magenta,      
    urlcolor=cyan,
    citecolor=violet,
}
\usepackage{cleveref}

\makeatletter
\g@addto@macro\bfseries{\boldmath}
\makeatother

\def \dif {\mathrm{d}}

\begin{document}

\title{Velocity decorrelation functions of high-energy cosmic rays propagating in magnetic fields}
\email{deligny@ijclab.in2p3.fr}

\author{Olivier Deligny}
\affiliation{Laboratoire de Physique des 2 Infinis Ir\`ene Joliot-Curie (IJCLab)\\
CNRS/IN2P3, Universit\'{e} Paris-Saclay, Orsay, France}
\date{\today}

\begin{abstract}

Diffusion tensor coefficients play a central role in describing cosmic-ray transport in various astrophysical environments permeated with magnetic fields, which are usually modeled as a fluctuating field on top of a mean field. In this article, a formal derivation of these coefficients is presented by means of the calculation of velocity decorrelation functions of particles. It relies mainly on expanding the 2-pt correlation function of the (fluctuating) magnetic field experienced by the particles between two successive times in the form of an infinite Dyson series and retaining a class of terms that converge to a physical solution. Subsequently, the velocity decorrelation functions, themselves expressed as Dyson series, are deduced from an iteration procedure that improves on the partial summation scheme. The results are shown to provide approximate solutions compared to those obtained by Monte-Carlo simulations as long as the Larmor radius of the particles is larger than at least one tenth of the largest scale of the turbulence.   

\end{abstract}

\maketitle

\section{Introduction} 
\label{sec:intro}

Cosmic rays compose less than one particle out of ten million in the interstellar gas. Still, their average energy density is similar to that of the gas. A small proportion of particles has therefore appropriated a substantial part of the available energy. The population of GeV cosmic rays is the most numerous and has important influence on the gas dynamics, heating and cooling the inter-stellar medium, launching large-scale outflows and modifying the phase structure of the background gas~\cite{Pfrommer:2017jau,Butsky_2018, Suoqing_2020,2022MNRAS.516.3470H}. Modeling their transport calls for highly non-linear processes, as they excite magnetic structures that control themselves the transport. When their streaming through the background plasma is faster than the Alfv\'en speed, cosmic rays transfer part of their energy and momentum to the gas by exciting instabilities~\cite{Kulsrud:1969zz,Mannheim:1994sv,Ensslin:2006mt,Guo:2007hv}. The streaming is possibly enhanced by the reduction of the Alfv\'en waves damped by interstellar dust grains~\cite{Squire_2021,Ji_2022} or by magnetrohydrodynamic turbulence cascade~\cite{2004ApJ...604..671F}. Recent studies have probed the  key role of those cosmic rays on driving strong galactic winds or on matching the observed ionization structure of the circum-galactic medium (the diffuse gas made of the majority of galactic baryons outside of the galactic disk)~\cite{Butsky_2018,Suoqing_2020,Quataert_2021}, or on star and galaxy formation~\cite{2022MNRAS.516.3470H}.

The energy density of trans-GeV cosmic rays does not dominate any longer that of the gas. Alfv\'en waves are then considered as an ``extrinsic turbulence'' excited by turbulent motions in the thermal gas~\cite{1966AnAp...29..645F,1971ApJ...170..265S}. The study presented in this article focuses on the transport of cosmic rays in this type of extrinsic turbulence. These environments constitute collisionless turbulent plasmas in which particles propagate or are accelerated. The confinement and transport of high-energy cosmic rays are governed by their scattering off the fluctuating magnetic fields, which act as an effective source of collisions~\cite{1972ApJ...172..319J}. Considering these collisions as a relaxation process, they tend to bring the average velocity distribution of the particles to its isotropic mean~\cite{PhysRev.94.511}. Under these conditions, the flux of particles can be related to their gradient of density by means of a diffusion tensor $D_{ij}$, which can be expressed in terms of the magnetic field unit vector $\mathbf{b}$, the diffusion coefficients parallel and perpendicular to the mean field $D_\parallel$ and $D_\perp$, and the anti-symmetric diffusion coefficient $D_A$ describing the particle drifts as~\cite{1990ApJ...361..162J}
\begin{equation}
\label{eqn:Dij}
    D_{ij}=D_\perp\delta_{ij}+\left(D_\parallel-D_\perp\right)b_ib_j+D_A\epsilon_{ijk}b_k.
\end{equation}
The determination of the coefficients proves to be a difficult task, as it requires dealing with a highly non-linear problem in several respects. Robust estimates rely therefore on numerical simulations exploring wide ranges of particle rigidities  and turbulence levels~\cite{1999ApJ...520..204G,PhysRevD.65.023002,Candia:2004yz,Parizot:2004wh,DeMarco:2007ux,Globus:2007bi,2010ApJ...711..997H,Plotnikov:2011me,Harari:2013pea,Fatuzzo:2014hua,2015ApJ...798...59S,Snodin:2015fza,Subedi:2016xwd,Giacinti:2017dgt,Reichherzer:2019dmb,Dundovic:2020sim,Reichherzer:2021yyd,Reichherzer:2023hae}. They show that the various approximations proposed in the literature fail to reproduce quantitatively and qualitatively the numerical results in various ranges of  turbulence levels~\cite{PhysRevD.65.023002,Fraschetti:2012cm}. Estimations based on the classical scattering theory~\cite{1975Ap&SS..32...77F}, on the quasi-linear theory~\cite{1966ApJ...146..480J,1973ApJ...183.1029J}, or on the intuitive ansatz using exponential decreases in the velocity decorrelation functions proposed in~\cite{1997ApJ...485..655B} do not capture the memory effects uncovered in simulations and yield to diffusion coefficients largely underestimated or overestimated. In this paper, a formal derivation based on first principles is explored. The range of application pertains to particles with a Larmor radius larger than at least one tenth of the largest scale of the turbulence. Consequently, the solution can be considered as an approximate one to describe the transport of cosmic rays in the high-rigidity regime (quasi-ballistic propagation) and in the rigidity range marking the transition between gyroresonant scattering (i.e. Larmor radius of particles entering in resonance with wavelengths of the turbulence) and quasi-ballistic propagation. 

The derivation builds on the theoretical one presented in~\cite{Plotnikov:2011me}. In that work, an estimation of the average velocity of the particles propagating in the turbulence as a function of time, expressed as a Dyson series, is achieved by using a white-noise model for the 2-pt function of the magnetic field experienced between two successive times. Such a modeling proves to be accurate in the high rigidity regime, in which the values of the magnetic field experienced by the particles decorrelate on time scales much smaller than that of the scattering. Under these conditions, the summation of the Dyson series gives rise to an exponential decay of the average velocity characteristic of a Markovian process. The diffusion coefficients are then inferred from velocity correlation function, $\langle v_{0i}v_{j}(t)\rangle$, through a time integration~\cite{Kubo:1957mj}, 
\begin{equation}
\label{eqn:Dij-Kubo}
    D_{ij}(t)=\int_0^{t}\dif t'~\langle v_{0i}v_{j}(t')\rangle,
\end{equation}
in the limit that $t\rightarrow\infty$. Here, $v_{0i}\equiv v_i(t=0)$ and $\langle\cdot\rangle$ stands for the average quantities, taken over several space and time correlation scales of the turbulent field. Since cosmic rays are high-energy relativistic particles, the norm of the velocity is identified to $c$ for convenience.

These results were extended to a range of  rigidities gyroresonant with the power spectrum of turbulence by allowing a short-time memory for the time evolution of the particle velocities through a red-noise approximation for the 2-pt function of the magnetic field experienced between two successive times~\cite{Deligny:2021wyi}. The extension was limited, however, to the case of pure turbulence. In this paper, the same kind of formalism is applied to any turbulence level. In Section~\ref{sec:transport}, the relevant Dyson series and diagrammatic representations are introduced to infer the time evolution of the particle velocities. The modeling, beyond white- or red-noise approximations, of the 2-pt function of the magnetic field experienced between two successive times is addressed in Section~\ref{sec:model-dbdb}. The general diagrammatic technique to carry out the partial-summation of the Dyson series for the velocity decorrelation functions is presented in Section~\ref{sec:summation}. The formalism is then applied to the case of parallel diffusion coefficient in Section~\ref{sec:parallel},  perpendicular diffusion coefficient in Section~\ref{sec:perpendicular}, and anti-symmetric diffusion coefficient in Section~\ref{sec:anti-symmetric}. Based on Monte-Carlo results (see Appendix~\ref{app:MC} for details about the Monte-Carlo generator), the limitations of the results obtained are underlined. General conclusions are drawn in Section~\ref{sec:discussion}. In addition, analytical approximations obtained in the red-noise approximation for the 2-pt function of the magnetic field experienced between two successive times are given in Appendix~\ref{app:red-noise}.

\section{Transport of cosmic rays in magnetic fields}
\label{sec:transport}

We are interested in determining the moments of $v_i(t)$ to derive a workable expression for Eqn.~\ref{eqn:Dij-Kubo}. Adopting the convention of implicit summation on repeated indices throughout the paper, the velocity of each test-particle is governed by the Lorentz-Newton equation of motion, 
\begin{equation}
\label{eqn:LorentzNewton}
    \dot{v}_i(t)=\delta\Omega ~\epsilon_{ijk}v_j(t)\delta b_k(t)+\Omega_0 ~\epsilon_{ijk}v_j(t) b_{0k}(t).
\end{equation}
Here, $\delta\Omega=c^2Z|e|\delta B/E$ is the gyrofrequency related to the turbulence with root mean square $\delta B$ for each component, $\Omega_0$ is the one related to the mean field oriented, to fix the ideas, such that $\mathbf{B}_0=B_0\mathbf{u}_z$, $Z|e|$ and $E$ the electric charge and the energy of the particle respectively, and $\delta b_k(t)\equiv\delta b_k(\mathbf{x}(t))$ the $k$-th component of the turbulence (expressed in units of $\delta B$) at the spatial coordinate $\mathbf{x}(t)$, which corresponds to the position of the test-particle at time $t$. A formal solution for $\langle v_{i}(t)\rangle$ can be obtained by expressing the solution of Eqn.~\ref{eqn:LorentzNewton} as an infinite number of Dyson series, each combining terms in powers of $\delta \mathbf{b}$ coupled to terms in powers of $\mathbf{B}_0$. Dealing with such an infinite number of Dyson series is however hardly manageable. To circumvent this difficulty, we use the auxiliary variable introduced in~\cite{Plotnikov:2011me}, $w_i(t)=R_{ij}(\Omega_0t)v_j(t)$, with $\hat{R}(\Omega_0t)$ the rotation matrix of angle $\Omega_0 t$ around $\mathbf{u}_z$. The equation of motion for $\mathbf{w}$ is then
\begin{multline}
    \label{eqn:LorentzNewton-w-1}
    R^{-1}_{ij}(\Omega_0t)\dot{w}_j(t)+\dot{R}^{-1}_{ij}(\Omega_0t)w_j(t)=\\
    \delta\Omega ~\epsilon_{ijk}R^{-1}_{j\ell}(\Omega_0t)w_\ell(t)\delta b_k(t)+\Omega_0 ~\epsilon_{ijk}R^{-1}_{j\ell}(\Omega_0t)w_\ell(t) b_{0k},
\end{multline}
which, taking advantage of $b_{0k}=\delta_{kz}$ throughout the paper, reads as
\begin{multline}
    \label{eqn:LorentzNewton-w-2}
    \dot{w}_i(t)=\delta\Omega ~R_{ij}(\Omega_0t)\epsilon_{jk\ell}R^{-1}_{km}(\Omega_0t)w_m(t)\delta b_\ell(t)\\
    -R_{ij}(\Omega_0t)\dot{R}^{-1}_{jm}(\Omega_0t)w_m(t)\\
    +\Omega_0 ~R_{ij}(\Omega_0t)\epsilon_{jkz}R^{-1}_{km}(\Omega_0t)w_m(t).    
\end{multline}
Noting that $\dot{R}^{-1}_{ij}(\Omega_0t)=\Omega_0\epsilon_{i\ell z}R^{-1}_{\ell j}(\Omega_0t)$, the two terms of the second line of Eqn.~\ref{eqn:LorentzNewton-w-2} cancel so that
\begin{equation}
    \label{eqn:LorentzNewton-w}
    \dot{w}_i(t)=\delta\Omega ~R_{ij}(\Omega_0t)\epsilon_{jk\ell}R^{-1}_{km}(\Omega_0t)w_m(t)\delta b_\ell(t).  
\end{equation}
Hence, the equation of motion for $\mathbf{w}$ is similar to that of $\mathbf{v}$ in a pure turbulence except for the action of the rotation matrices: for each infinitesimal time step, $\mathbf{w}$ is rotated by $-\Omega_0t$ around $\mathbf{u}_z$ prior to undergoing the impact of $\delta\mathbf{b}$, and is rotated by $+\Omega_0t$ afterwards. 

\begin{widetext}
The formal solution for the average $\langle w_{i_0}(t)\rangle$ can be expressed as a \textit{single} Dyson series: 
\begin{multline}
    \label{eqn:dyson}
    \langle w_{i_0}(t)\rangle = w_{0i_0}+\sum_{p=1}^\infty \delta\Omega^p ~ \epsilon_{k_1m_1n_1}\epsilon_{k_2m_2n_2}\dots\epsilon_{k_pm_pn_p} w_{0i_p} \int_0^t \dif t_1\int_0^{t_1}\dif t_2\dots\int_0^{t_{p-1}}\dif t_p  \\ 
    R_{i_0k_1}(\Omega_0t_1)R_{i_1k_2}(\Omega_0t_2)\cdots R_{i_{p-1}k_p}(\Omega_0t_p)R^{-1}_{m_1i_1}(\Omega_0t_1)R^{-1}_{m_2i_2}(\Omega_0t_2)\cdots R^{-1}_{m_pi_p}(t_p)\langle\delta b_{n_1}(t_1)\dots\delta b_{n_p}(t_p)\rangle,
\end{multline} 
\end{widetext}
using the shortcut notation $w_{0i_0}\equiv w_{i_0}(t=0)$. In the right hand side of this expression, the expectation value $\langle\delta b_{n_1}(t_1)\dots\delta b_{n_p}(t_p)\rangle$ can be related, in the Gaussian regime, to all permutations of products of contraction of pairs by using the Wick theorem,
\begin{equation}
    \label{eqn:wick}
    \langle\delta b_{n_1}(t_1)\dots\delta b_{n_p}(t_p)\rangle=\sum_{\mathrm{pairings}}^{}\prod_{j<\ell} \langle\delta b_{n_j}(t_j)\delta b_{n_\ell}(t_\ell)\rangle,
\end{equation}
where the notation $\sum_{\mathrm{pairings}}^{}\prod_{j<\ell} $ stands for the $(2n_p-1)!!$ possible permutations of pairs with $t_j<t_\ell$. Without loss of generality, we consider in this study the case of a 3D isotropic turbulence without helicity. The 2-pt function is then dependent on the time difference only, 
\begin{equation}
    \label{eqn:dbdb-iso3D}
    \langle\delta b_{n_j}(t_j)\delta b_{n_\ell}(t_\ell)\rangle=\frac{\varphi(t_j-t_\ell)}{3}\delta_{n_jn_\ell},
\end{equation}
with $\varphi(t)$ a function that describes the correlation of the turbulence experienced by a test-particle along its path at two different times. An expression for $\varphi(t)$, inferred from a formal derivation going beyond the white-noise approximation or the red-noise one (see Appendix~\ref{app:red-noise}), will be presented in Section~\ref{sec:model-dbdb}. 

\begin{widetext}
On inserting Eqn.~\ref{eqn:wick} and Eqn.~\ref{eqn:dbdb-iso3D} into Eqn.~\ref{eqn:dyson}, the Dyson series reads as
\begin{multline}
    \label{eqn:dyson-bis}
    \langle w_{i_0}(t) \rangle = w_{0i_0}+\sum_{p=1}^\infty \left(\frac{\delta\Omega^2}{3}\right)^p w_{0i_{2p}} \int_0^t \dif t_1\int_0^{t_1}\dif t_2\dots\int_0^{t_{2p-1}}\dif t_{2p}
    \sum_{\mathrm{pairings}}^{}\prod_{j<\ell}  \\
    \left(R_{i_{j-1}k_j}(\Omega_0t_j)R_{i_{\ell-1}k_j}(\Omega_0t_\ell)R^{-1}_{m_ji_j}(\Omega_0t_j)R^{-1}_{m_ji_\ell}(\Omega_0t_\ell)-R_{i_{j-1}k_j}(\Omega_0t_j)R_{i_{\ell-1}k_\ell}(\Omega_0t_\ell)R^{-1}_{k_\ell i_j}(\Omega_0t_j)R^{-1}_{k_ji_\ell}(\Omega_0t_\ell)\right)\varphi(t_j-t_\ell),
\end{multline} 
which is the relevant equation to determine the time evolution of the auxiliary variable $\mathbf{w}(t)$ and subsequently of the particle velocity $\mathbf{v}(t)=\hat{R}^{-1}(\Omega_0t)\mathbf{w}(t)$. The various terms of the expansion can be conveniently represented using diagrammatic rules~\cite{Bourret1962,1966AnAp...29..645F}. In the following, we denote the ``mass propagator'' function $w(t)$, defined such that $w_i(t)=w(t)\hat{u}_i$, by a double line, while a single line stands for the corresponding ``free propagator'' corresponding to $w^{(0)}(t)=1$. On the other hand, considering a contraction of a pair as an ``interaction'' in which a free propagator is inserted, $\langle \delta b_{i_1}(t_{j_1})\delta b_{i_2}(t_{j_2})\rangle = \langle \delta b_{i_1}(t_{j_1})u^{(0)}(t)\delta b_{i_2}(t_{j_2})\rangle$, a curved dotted line connecting two ``vertices'' then stands for a time-ordered integration over an average product of two stochastic fields. In this manner, the first term of the summation ($p=1$) is generically represented as
\begin{equation}\label{eqn:diagram-2}
\begin{tikzpicture}
\node at (2.4,-0.3) {\( t,i_0 \)};
\node at (1.7,-0.3) {\( t_1 \)};
\node at (0.7,-0.3) {\( t_2 \)};
\node at (0.,-0.3) {\( 0,i_2 \)};
\node at (8.0,0.05) {\( =~\displaystyle \frac{\delta\Omega^2}{3}\displaystyle\int_0^t\dif t_1\int_0^{t_1}\dif t_2~\bigg(R_{i_0k_1}(\Omega_0t_1)R_{i_1k_1}(\Omega_0t_2)R^{-1}_{m_1i_1}(\Omega_0t_1)R^{-1}_{m_1i_2}(\Omega_0t_2) \)};
\node at (9.8,-1) {\( -\displaystyle R_{i_0k_1}(\Omega_0t_1)R_{i_1k_2}(\Omega_0t_2)R^{-1}_{k_2 i_1}(\Omega_0t_1)R^{-1}_{k_1i_2}(\Omega_0t_2)\bigg)\varphi(t_1-t_2), \)};
\begin{feynman}
    \vertex (a1);
    \vertex [right=0.7cm of a1] (a2);
    \vertex [right=1.0cm of a2] (a3);
    \vertex [right=0.7cm of a3] (a4);
    \diagram*{
      (a1) --[plain] (a4),  
      (a2) -- [scalar, out=90, in=90, looseness=2.0,thick] (a3),
    };
  \end{feynman}
\end{tikzpicture}
\end{equation}
which, using the expression of the rotation matrices, reads explicitly as 
\begin{equation}\label{eqn:diagram-2-zz}
\begin{tikzpicture}
\node at (2.4,-0.3) {\( t,z \)};
\node at (1.7,-0.3) {\( t_1 \)};
\node at (0.7,-0.3) {\( t_2 \)};
\node at (0.,-0.3) {\( 0,z \)};
\node at (6.4,0) {\( =~\displaystyle -2\frac{ \delta\Omega^2}{3}\displaystyle\int_0^t\dif t_1\int_0^{t_1}\dif t_2~\cos{\Omega_0(t_1-t_2)}\varphi(t_1-t_2), \)};
\begin{feynman}
    \vertex (a1);
    \vertex [right=0.7cm of a1] (a2);
    \vertex [right=1.0cm of a2] (a3);
    \vertex [right=0.7cm of a3] (a4);
    \diagram*{
      (a1) --[plain] (a4),  
      (a2) -- [scalar, out=90, in=90, looseness=2.0,thick] (a3),
    };
  \end{feynman}
\end{tikzpicture}
\end{equation}
in the case $i_0=i_2=z$. For $i_0=i_2=x$, on the other hand, the contraction of the rotation matrices leads to
\begin{equation}\label{eqn:diagram-2-xx}
\begin{tikzpicture}
\node at (2.4,-0.3) {\( t,x \)};
\node at (1.7,-0.3) {\( t_1 \)};
\node at (0.7,-0.3) {\( t_2 \)};
\node at (0.,-0.3) {\( 0,x \)};
\node at (6.7,0) {\( =~\displaystyle -\frac{\delta\Omega^2}{3}\displaystyle\int_0^t\dif t_1\int_0^{t_1}\dif t_2~(1+\cos{\Omega_0(t_1-t_2)})\varphi(t_1-t_2). \)};
\begin{feynman}
    \vertex (a1);
    \vertex [right=0.7cm of a1] (a2);
    \vertex [right=1.0cm of a2] (a3);
    \vertex [right=0.7cm of a3] (a4);
    \diagram*{
      (a1) --[plain] (a4),  
      (a2) -- [scalar, out=90, in=90, looseness=2.0,thick] (a3),
    };
  \end{feynman}
\end{tikzpicture}
\end{equation}
It will prove useful to express diagrams in the Laplace reciprocal space. In that case, indices referring to initial and final times $0$ and $t$ are removed. The generic change of variables $t=x+x_1+\cdots +x_p$, $t_1=x_1+\cdots +x_p$, $t_2=x_2+\cdots +x_p$,$\cdots$, $t_p=x_p$ allows for sending all integration boundaries between 0 and $+\infty$ for the $x_i$ variables. In this manner, the diagram of Eqn.~\ref{eqn:diagram-2-zz} for instance becomes
\begin{equation}\label{eqn:diagram-2-zz-reciprocal}
\begin{tikzpicture}
\node at (2.4,-0.3) {\( z \)};
\node at (1.7,-0.3) {\( t_1 \)};
\node at (0.7,-0.3) {\( t_2 \)};
\node at (0.,-0.3) {\( z \)};
\node at (7.7,0.05) {\( =~ \displaystyle -2\frac{\delta\Omega^2}{3}\displaystyle\int_0^\infty\dif x~\mathrm{e}^{-sx}\int_0^\infty\dif x_2~\mathrm{e}^{-sx_2}\int_0^\infty\dif x_1~\mathrm{e}^{-sx_1}\cos{\Omega_0(x_1)}\varphi(x_1) \)};
\node at (5.7,-1) {\( =~ \displaystyle -2\frac{\delta\Omega^2}{3}\mathcal{L}^2[1](s)\mathcal{L}[\varphi(x)\cos{\Omega_0x}](s), \)};
\begin{feynman}
    \vertex (a1);
    \vertex [right=0.7cm of a1] (a2);
    \vertex [right=1.0cm of a2] (a3);
    \vertex [right=0.7cm of a3] (a4);
    \diagram*{
      (a1) --[plain] (a4),  
      (a2) -- [scalar, out=90, in=90, looseness=2.0,thick] (a3),
    };
  \end{feynman}
\end{tikzpicture}
\end{equation}
with $\mathcal{L}[f(x)](s)$ the Laplace transform of $f(x)$ expressed as a function of the variable $s$. For $p=2$, in addition to the unconnected two-loop contribution, there are nested and crossed diagrams. In the case $i_0=i_2=z$ for instance, the contribution of the nested diagram is
\begin{equation}\label{eqn:diagram-4n-zz}
\begin{tikzpicture}
\node at (3.1,-0.3) {\( t,z \)};
\node at (2.6,-0.3) {\( t_1 \)};
\node at (1.9,-0.3) {\( t_2 \)};
\node at (1.2,-0.3) {\( t_3 \)};
\node at (0.6,-0.3) {\( t_4 \)};
\node at (0.,-0.3) {\( 0,z \)};
\node at (10.8,0.05) {\( =\displaystyle ~4\left(\frac{\delta\Omega^2}{3}\right)^2\displaystyle\int_0^t\dif t_1\int_0^{t_1}\dif t_2\int_0^{t_2}\dif t_3\int_0^{t_3}\dif t_4~\cos{\Omega_0\left(\frac{t_2-t_3}{2}\right)}\cos{\Omega_0\left(t_1-\frac{t_2-t_3}{2}-t_4\right)}\varphi(t_1-t_4)\varphi(t_2-t_3), \)};
\begin{feynman}
    \vertex (a1);
    \vertex [right=0.5cm of a1] (a2);
    \vertex [right=0.7cm of a2] (a3);
    \vertex [right=0.7cm of a3] (a4);
    \vertex [right=0.7cm of a4] (a5);
    \vertex [right=0.5cm of a5] (a6);
    \diagram*{
      (a1) --[plain] (a6),  
      (a2) -- [scalar, out=90, in=90, looseness=1.5,thick] (a5),
      (a3) -- [scalar, out=90, in=90, looseness=2.0,thick] (a4),
    };
\end{feynman}
\end{tikzpicture}
\end{equation}
while that of the crossed diagram is
\begin{equation}\label{eqn:diagram-4c-zz}
\begin{tikzpicture}
\node at (3.1,-0.3) {\( t,z \)};
\node at (2.6,-0.3) {\( t_1 \)};
\node at (1.9,-0.3) {\( t_2 \)};
\node at (1.2,-0.3) {\( t_3 \)};
\node at (0.6,-0.3) {\( t_4 \)};
\node at (0.,-0.3) {\( 0,z \)};
\node at (9.5,0.05) {\( \,\,=\displaystyle ~2\left(\frac{\delta\Omega^2}{3}\right)^2\displaystyle\int_0^t\dif t_1\int_0^{t_1}\dif t_2\int_0^{t_2}\dif t_3\int_0^{t_3}\dif t_4~\cos{\Omega_0\left(t_1-t_2-t_3+t_4\right)}\varphi(t_1-t_3)\varphi(t_2-t_4), \)};
\begin{feynman}
    \vertex (a1);
    \vertex [right=0.5cm of a1] (a2);
    \vertex [right=0.7cm of a2] (a3);
    \vertex [right=0.7cm of a3] (a4);
    \vertex [right=0.7cm of a4] (a5);
    \vertex [right=0.5cm of a5] (a6);
    \diagram*{
      (a1) --[plain] (a6),  
      (a2) -- [scalar, out=90, in=90, looseness=1.5,thick] (a4),
      (a3) -- [scalar, out=90, in=90, looseness=1.5,thick] (a5),
    };
\end{feynman}
\end{tikzpicture}
\end{equation}
and so on and so forth for higher values of $p$ and other values of $i_{2p}$ and $i_0$ (see Appendix~\ref{app:p=2} for $p=2$ contributions with other configurations of $i_0$ and $i_{2p}$). In these conditions, the mass propagator results from the infinite summation
\begin{equation}\label{eqn:udyson-ex}
\begin{tikzpicture}
\node at (15.0,-0.3) {\( t,j \)};
\node at (12.5,-0.3) {\( 0,i \)};  \node at (11.8,-0.3) {\( t,j \)};
\node at (9.3,-0.3) {\( 0,i \)};  \node at (8.6,-0.3) {\( t,j \)};
\node at (5.7,-0.3) {\( 0,i \)};  \node at (5.0,-0.3) {\( t,j \)};
\node at (3.3,-0.3) {\( 0,i \)};  \node at (2.6,-0.3) {\( t,j \)};
\node at (1.7,-0.3) {\( 0,i \)};  \node at (0.9,-0.3) {\( t,j \)};
\node at (0.0,-0.3) {\( 0,i \)};  \begin{feynman}
    \vertex (a1);
    \vertex [right=1.0cm of a1] (a2){~=~};
    \vertex [right=0.4cm of a2] (a3);
    \vertex [right=1.0cm of a3] (a4){~+~};
    \vertex [right=0.4cm of a4] (a5);
    \vertex [right=0.5cm of a5] (a6);
    \vertex [right=0.7cm of a6] (a7);
    \vertex [right=0.5cm of a7] (a8){~+~};
    \vertex [right=0.4cm of a8] (a9);
    \vertex [right=0.5cm of a9] (a10);
    \vertex [right=0.7cm of a10] (a11);
    \vertex [right=0.5cm of a11] (a12);
    \vertex [right=0.7cm of a12] (a13);
    \vertex [right=0.5cm of a13] (a14){~+~};
    \vertex [right=0.4cm of a14] (a15);
    \vertex [right=0.5cm of a15] (a16);
    \vertex [right=0.5cm of a16] (a17);
    \vertex [right=0.5cm of a17] (a18);
    \vertex [right=0.5cm of a18] (a19);
    \vertex [right=0.5cm of a19] (a20){~+~};
    \vertex [right=0.4cm of a20] (a22);
    \vertex [right=0.5cm of a22] (a23);
    \vertex [right=0.5cm of a23] (a24);
    \vertex [right=0.5cm of a24] (a25);
    \vertex [right=0.5cm of a25] (a26);
    \vertex [right=0.5cm of a26] (a27){~+~...~,};
    \diagram*{
      (a1) --[double,thick] (a2),
      (a3) --[plain] (a4),
      (a5) --[plain] (a8),  
      (a6) -- [scalar, out=90, in=90, looseness=2.0,thick] (a7),
      (a9) --[plain] (a14),  
      (a10) -- [scalar, out=90, in=90, looseness=2.0,thick] (a11),
      (a12) -- [scalar, out=90, in=90, looseness=2.0,thick] (a13),
      (a15) --[plain] (a20),  
      (a16) -- [scalar, out=90, in=90, looseness=1.5,thick] (a18),
      (a17) -- [scalar, out=90, in=90, looseness=1.5,thick] (a19),
      (a22) --[plain] (a27),  
      (a23) -- [scalar, out=90, in=90, looseness=1.5,thick] (a26),
      (a24) -- [scalar, out=90, in=90, looseness=2.0,thick] (a25),
    };
  \end{feynman}
\end{tikzpicture}
\end{equation}
which can be symbolically summarized by introducing a blob that stands for the sum over all the possible connected diagrams as
\begin{equation}\label{eqn:udyson-bis}
\begin{tikzpicture}
\node at (6.4,-0.3) {\( t,j \)};
\node at (3.8,-0.3) {\( 0,i \)};    \node at (3.1,-0.3) {\( t,j \)};
\node at (1.9,-0.3) {\( 0,i \)};    \node at (1.2,-0.3) {\( t,j \)};
\node at (0.0,-0.3) {\( 0,i \)};    \begin{feynman}
    \vertex (a1);
    \vertex [right=1.2cm of a1] (a2){~=~};
    \vertex [right=0.4cm of a2] (a3);
    \vertex [right=1.2cm of a3] (a4){~+~};
    \vertex [right=0.4cm of a4] (a5);
    \vertex [right=1.0cm of a5] (a6);
    \node[right=0cm of a6,blob] (a7);
    \vertex [right=1.5cm of a7] (a8){.};
    \diagram*{
      (a1) --[double,thick] (a2),
      (a3) --[plain] (a4),
      (a5) --[plain] (a6),        
      (a7) --[double,thick]  (a8)
    };
  \end{feynman}
\end{tikzpicture}
\end{equation}

\end{widetext}

\section{Partial-summation approximation of the 2-pt function $\langle\delta b_{n_j}(t_j)\delta b_{n_\ell}(t_\ell) \rangle$}
\label{sec:model-dbdb}

One key ingredient in the calculation of Eqn.~\ref{eqn:udyson-bis} is the 2-pt function of the fluctuating magnetic field $\langle\delta b_{n_j}(t_j)\delta b_{n_\ell}(t_\ell) \rangle$ experienced by a test particle. Based on Eqn.~\ref{eqn:dbdb-iso3D}, we focus in this section on deriving an expression for the function $\varphi(t)$. 

The fluctuating magnetic field, considered as a Gaussian random field with zero mean and $\langle\delta B_i(\mathbf{x})\delta B^\star_j(\mathbf{x'})\rangle =\delta B^2\delta_{ij}$, is denoted as $\delta\mathbf{B}(\mathbf{k})$ in the reciprocal Fourier space.  For an homogeneous turbulence, the 2-pt correlation function between two components of $\delta\mathbf{B}(\mathbf{x})$ is invariant under spatial translations. This implies that two Fourier components of the field are uncorrelated at different wavenumber vectors: 
\begin{equation}
\label{eqn:dBdB}
\langle\delta B_i(\mathbf{k})\delta B^\star_j(\mathbf{k'})\rangle=P_{ij}(\mathbf{k})\delta(\mathbf{k}-\mathbf{k'}),
\end{equation}
where $P_{ij}$ is the spectral tensor defined as the Fourier transform of the 2-pt correlation function. Combining the solenoidal nature of the field, its isotropy and its absence of helicity, $P_{ij}$ takes the form~\cite{Batchelor70}
\begin{equation}
\label{eqn:Pij}
P_{ij}(\mathbf{k})=\frac{\mathcal{E}(k)}{4\pi k^2}\left(\delta_{ij}-\frac{k_ik_j}{k^2}\right),
\end{equation}
with $\mathcal{E}(k)$ the kinetic energy spectrum of the turbulence, which, for a Kolmogorov turbulence, is defined between $k_\mathrm{min}$ and $k_\mathrm{max}$ as
\begin{equation}
\label{eqn:Ek}
\mathcal{E}(k)=\frac{(2\pi)^{2/3}\delta B^2}{3\left(L_{\mathrm{max}}^{2/3}-L_{\mathrm{min}}^{2/3}\right)}k^{-5/3}.
\end{equation}
The minimum wavenumber vector $\mathbf{k}_\mathrm{min}$ is related to the distance $L_{\mathrm{max}}$ over which the correlation function is non-zero (size of the largest ``eddies''), while the maximum one $\mathbf{k}_\mathrm{max}$ is related to the scale $L_{\mathrm{min}}$ at which the dissipation rate of the turbulence overcomes the energy cascade rate.

With these ingredients, the 2-pt function of the fluctuating magnetic field $\langle\delta b_{n_j}(t_j)\delta b_{n_\ell}(t_\ell) \rangle$ experienced by a test particle can be expressed as 
\begin{equation}
    \langle \delta b_{i}(t)\delta b_{j}(0)\rangle\simeq \int_{\mathbf{k}_\mathrm{min}}^{\mathbf{k}_\mathrm{max}} \dif\mathbf{k}\frac{\mathcal{E}(k)}{4\pi k^2}\left(\delta_{ij}-\frac{k_ik_j}{k^2}\right)\langle \mathrm{e}^{\mathrm{i}\mathbf{k}\cdot\mathbf{x}(t)}\rangle,
\end{equation}
where the Corrsin approximation has been used~\cite{Corrsin59}. Different approximations have been proposed to estimate the factor $\langle e^{\mathrm{i}\mathbf{k}\cdot\mathbf{x}(t)}\rangle$~\cite{1966ApJ...146..480J,1973ApJ...183.1029J,PhysRevD.65.023002}. In this study, to evaluate it, we start from the formal expansion 
\begin{equation}
    \langle \mathrm{e}^{\mathrm{i}\mathbf{k}\cdot\mathbf{x}(t)}\rangle=\sum_{n\geq 0} \frac{\mathrm{i}^n}{n!}\int_0^t \dif t_1~\dots\int_0^t \dif t_n\langle(\mathbf{k}\cdot\mathbf{v}(t_1))\dots (\mathbf{k}\cdot\mathbf{v}(t_n))\rangle, 
\end{equation}
where the substitution $\mathbf{x}(t)=\int_0^t~dt'~\mathbf{v}(t')$ has been used. Next, the $n$-point correlation function entering into the integrand expression can be substituted for the sum of all possible contraction of pairs (Wick theorem). Because the process that draws at random the wavenumber vectors of the turbulence is independent from that governing the velocity decorrelation of the test particles, 
each pair can be approximated as
\begin{equation}
    \langle(\mathbf{k}\cdot\mathbf{v}(t_1)) (\mathbf{k}\cdot\mathbf{v}(t_2))\rangle \simeq (kc)^2 \langle \cos{\mathbf{\hat{k}} \cdot\mathbf{\hat{v}}(t_1)}\cos{\mathbf{\hat{k}} \cdot\mathbf{\hat{v}}(t_2)}\rangle.
\end{equation}
The arguments of the cosines are the pitch angles between the particle velocities and the wavenumber vectors of the turbulence. The decorrelation of the pitch angle is then assumed to decay exponentially,
\begin{equation}
    \langle(\mathbf{k}\cdot\mathbf{v}(t_1)) (\mathbf{k}\cdot\mathbf{v}(t_2))\rangle \simeq (kc)^2\mathrm{e}^{-(t_1-t_2)/\xi(k)},
\end{equation}
expression that requires the introduction of a correlation time scale $\xi$. Guided by Monte-Carlo simulations that show a longer falloff timescale for the 2-pt
function of the fluctuating magnetic field for large $k$ compared to small $k$, a dependency in $(kc)^{-1}$ turns out to reproduce the main features of $\varphi(t)$ for a reduced rigidity (Larmor radius conventionally related to the turbulence only and expressed in units of the largest eddy scale $L_\mathrm{max}$) $\rho=1$. An additional dependency in $\rho$ is introduced through $\xi(k,\rho)=A\rho^B/(kc)$; $A\simeq 1$ and $B\simeq 0.5$ are found to provide a good compromise to cover the gyroresonant and high-rigidity regimes. Some algebra then leads to
\begin{multline}
\label{eqn:dbdb}
    \langle \mathrm{e}^{\mathrm{i}\mathbf{k}\cdot\mathbf{x}(t)}\rangle\simeq \sum_{p\geq 0} (-(kc)^2)^p \int_0^t \dif t_1\int_0^{t_1} \dif t_2~\dots\int_0^{t_{2p-1}} \dif t_{2p} \\
    \sum_{\mathrm{pairings}}\prod_{\mathrm{pairs}~i<j}\mathrm{e}^{-(t_i-t_j)/\xi(k)}.
\end{multline}
To evaluate the right hand side, only pairs with $j=i+1$ are retained. Under this approximation, which corresponds to summing unconnected diagrams~\cite{Bourret1962}, our estimate of $\langle \mathrm{e}^{\mathrm{i}\mathbf{k}\cdot\mathbf{x}(t)}\rangle$, denoted with a subscript $0$, can be written in a compact non-linear manner:
\begin{equation}
\label{eqn:dbdb-bourret}
    \langle \mathrm{e}^{\mathrm{i}\mathbf{k}\cdot\mathbf{x}(t)}\rangle_0\simeq 1-(kc)^2\int_0^t\dif t_1\int_0^{t_1}\dif t_2~\mathrm{e}^{-\frac{t_1-t_2}{\xi(k)}}\langle \mathrm{e}^{\mathrm{i}\mathbf{k}\cdot\mathbf{x}(t-t_1)}\rangle_0.
\end{equation}
In the Laplace reciprocal space, the equation is then linear in $\mathcal{L}[\langle \mathrm{e}^{\mathrm{i}\mathbf{k}\cdot\mathbf{x}(t)}\rangle_0](s)$, which reads as
\begin{equation}
\label{eqn:dbdb-laplace}
    \mathcal{L}[\langle \mathrm{e}^{\mathrm{i}\mathbf{k}\cdot\mathbf{x}(t)}\rangle_0](s)=\frac{1+s\xi(k)}{(1+s\xi(k))s+(kc)^2\xi(k)},
\end{equation}
so that $\langle \mathrm{e}^{\mathrm{i}\mathbf{k}\cdot\mathbf{x}(t)}\rangle_0$ can be inferred from a numerical inverse Laplace transformation. Subsequently, we infer the partial-summation approximation for the 2-pt function of the fluctuating magnetic field experienced by a test particle as
\begin{multline}
\label{eqn:dbdb-final}
    \langle \delta b_{i}(t)\delta b_{j}(0)\rangle\simeq \frac{2\delta_{ij}}{3}\frac{(2\pi)^{2/3}\delta B^2}{3\left(L_{\mathrm{max}}^{2/3}-L_{\mathrm{min}}^{2/3}\right)}\\
    \int_{k_{\mathrm{min}}}^{k_{\mathrm{max}}}\dif k~k^{-5/3}\mathcal{L}^{-1}\bigg[\frac{1+s\xi(k)}{(1+s\xi(k))s+(kc)^2\xi(k)}\bigg](t),
\end{multline}
expression from which the function $\varphi(t)$ is deduced by identification with Eqn.~\ref{eqn:dbdb-iso3D}.

\begin{figure}[t]
\centering
\includegraphics[width=\columnwidth]{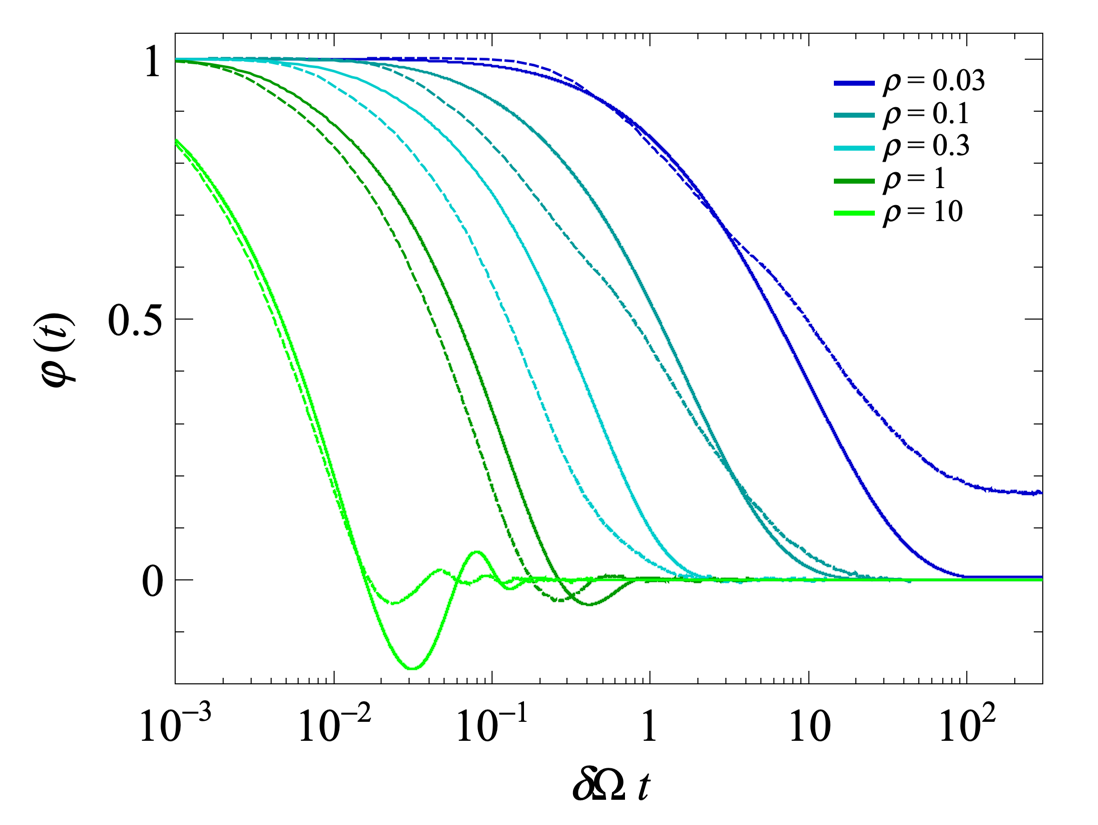}
\caption{2-pt correlation function of the magnetic field experienced by a test-particle as a function of the gyroperiod time scale $\delta\Omega t$ for five different rigidities. The dotted lines are from Monte-Carlo simulations.}
\label{fig:dbdb}
\end{figure}

As benchmark values typical of those relevant for the propagation of high-energy cosmic rays in the Galaxy, we use hereafter $\delta B=1~\upmu$G, $L_\mathrm{max}/L_\mathrm{min}=100$ and $L_\mathrm{max}=100~$pc. Examples of $\varphi(t)$ obtained from Eqn.~\ref{eqn:dbdb-final} are shown as a function of the gyroperiod scale $\delta\Omega~t$ as continuous lines in Fig.~\ref{fig:dbdb} for different reduced rigidities, while the corresponding results from Monte-Carlo simulations, taken from~\cite{Deligny:2023obh}, are displayed as the dotted lines. Results are displayed on a logarithmic scale for $\delta\Omega~t$ to appreciate fine similarities and differences between the model and the Monte-Carlo solution. As the rigidity increases, the time scale of correlation is observed to decrease significantly in terms of $\delta\Omega t$. Overall, the main features uncovered by the Monte-Carlo calculation are qualitatively well reproduced by the model, especially at high rigidities (in particular the oscillations around 0 that are absent in the quasi-linear theory or the red-noise approximation). However, some differences are observed quantitatively. The descent of the modeled $\varphi(t)$ with $\delta\Omega~t$ is delayed for $0.1\lesssim\rho\lesssim 1$ while, more importantly, it becomes too fast for $\rho\lesssim 0.1$ and unable to capture the slow decrease at large $\delta\Omega t$. Consequently, the use of the modeled $\varphi(t)$ function is limited to the rigidity range such that $\rho \gtrsim 0.1$.

\section{Summation scheme}
\label{sec:summation}

To carry out a summation of the Dyson series (Eqn.~\ref{eqn:dyson-bis} and Eqn.~\ref{eqn:udyson-bis}), we resort to a two-step iteration procedure. In the first iteration, we calculate the propagator that would be accurate in the case of a 2-pt function $\varphi(t)$ approximated by a Dirac function. As mentioned in the introduction, this approximation proves to be accurate in the high-rigidity regime in pure turbulence. It corresponds to the summation of the class of unconnected diagrams and represents the simplest partial summation scheme of the Dyson series~\cite{Bourret1962}. However, this scheme leads to a non-physical solution in the case of a non-zero mean field and in the gyroresonant regime. A more physical solution is then obtained by including classes of nested and crossed diagrams in the partial summation of the Dyson series, in which the first iteration of the propagator is inserted into each ordered time. We now detail this scheme below.

The propagator obtained by retaining only unconnected diagrams is denoted as $\langle w_i(t) \rangle_0$. For the sake of clarity, we distinguish its symbol from that of the final propagator by marking it with zigzags rather than a double line. This propagator is the solution to the equation
\begin{equation} \label{eqn:bourret}
\begin{tikzpicture}
\node at (5.8,-0.3) {\( t,i \)};
\node at (5.1,-0.3) {\( t_1 \)};
\node at (4.1,-0.3) {\( t_2 \)};
\node at (3.4,-0.3) {\( 0,j \)};
\node at (2.7,-0.3) {\( t,i \)};
\node at (1.7,-0.3) {\( 0,j \)};
\node at (1.0,-0.3) {\( t,i \)};
\node at (0.,-0.3) {\( 0,j \)};
\begin{feynman}
    \vertex (a1);
    \vertex [right=1.0cm of a1] (a2){~=~};
    \vertex [right=0.4cm of a2] (a3);
    \vertex [right=1.0cm of a3] (a4){~+~};
    \vertex [right=0.4cm of a4] (a5);
    \vertex [right=0.7cm of a5] (a6);
    \vertex [right=1.0cm of a6] (a7);
    \vertex [right=0.7cm of a7] (a8){.};

    \diagram*{
      (a1) --[boson] (a2),  
      (a3) --[plain] (a4),  
      (a5) --[plain] (a7),
      (a7) --[boson] (a8),  
      (a6) -- [scalar, out=90, in=90, looseness=2.0,thick] (a7),
    };
  \end{feynman}
\end{tikzpicture}
\end{equation}
\begin{widetext}
In view of the rules detailed in Section~\ref{sec:transport} governing the calculation of diagrams, the second term of the equation reads as
\begin{equation}\label{eqn:bourret-2}
\begin{tikzpicture}
\node at (2.4,-0.3) {\( t,i \)};
\node at (1.7,-0.3) {\( t_1 \)};
\node at (0.8,-0.3) {\( t_2 \)};
\node at (0.,-0.3) {\( 0,j \)};
\node at (10.1,0) {\( = \displaystyle \frac{\delta\Omega^2}{3}\displaystyle\int_0^t\dif t_1\int_0^{t_1}\dif t_2~\left(R^{-1}_{ij}(2\Omega_0(t_1-t_2)) -(1+2\cos\Omega_0(t_1-t_2))R^{-1}_{ij}(\Omega_0(t_1-t_2)\right) \varphi(t_1-t_2) \langle w_i(t-t_1)\rangle_0, \)};
\begin{feynman}
    \vertex (a1);
    \vertex [right=0.7cm of a1] (a2);
    \vertex [right=1cm of a2] (a3);
    \vertex [right=0.7cm of a3] (a4);
    \diagram*{
      (a1) --[plain] (a2),  
      (a2) --[plain] (a3),  
      (a3) --[boson] (a4),  
      (a2) -- [scalar, out=90, in=90, looseness=2.0,thick] (a3),
    };
  \end{feynman}
\end{tikzpicture}
\end{equation}
which gives rise to a linear term for $\mathcal{L}[\langle w_i(x)\rangle_0](s)$ in the Laplace reciprocal space:
\begin{equation}\label{eqn:bourret-2-laplace}
\begin{tikzpicture}
\node at (2.4,-0.3) {\( i \)};
\node at (1.7,-0.3) {\( t_1 \)};
\node at (0.8,-0.3) {\( t_2 \)};
\node at (0.,-0.3) {\( j \)};
\node at (8.5,0.05) {\( = \displaystyle \frac{\delta\Omega^2}{3}\mathcal{L}[\langle w_i(x)\rangle_0](s)\mathcal{L}[1](s)
\mathcal{L}\left[\left(R^{-1}_{ij}(2\Omega_0x) -(1+2\cos\Omega_0x)R^{-1}_{ij}(\Omega_0x)\right)\varphi(x)\right](s). \)};
\begin{feynman}
    \vertex (a1);
    \vertex [right=0.7cm of a1] (a2);
    \vertex [right=1cm of a2] (a3);
    \vertex [right=0.7cm of a3] (a4);
    \diagram*{
      (a1) --[plain] (a2),  
      (a2) --[plain] (a3),  
      (a3) --[boson] (a4),  
      (a2) -- [scalar, out=90, in=90, looseness=2.0,thick] (a3),
    };
  \end{feynman}
\end{tikzpicture}
\end{equation}
The propagator $\langle w_i(t)\rangle_0$ is therefore obtained through a numerical inverse of Laplace transform. Throughout this study, the Stehfest algorithm is used, with Stehfest number $N=20$~\cite{Stehfest}. 

The second summation scheme accounts, in addition to unconnected diagrams, for contributions from nested and crossed diagrams:
\begin{equation}\label{eqn:iter}
\begin{tikzpicture}
\node at (10.0,-0.3) {\( t,i \)};
\node at (9.3,-0.3) {\( t_1 \)};
\node at (8.6,-0.3) {\( t_2 \)};
\node at (7.9,-0.3) {\( t_3 \)};
\node at (7.2,-0.3) {\( t_4 \)};
\node at (6.5,-0.3) {\( 0,j \)};
\node at (5.8,-0.3) {\( t,i \)};
\node at (5.1,-0.3) {\( t_1 \)};
\node at (4.1,-0.3) {\( t_2 \)};
\node at (3.4,-0.3) {\( 0,j \)};
\node at (2.7,-0.3) {\( t,i \)};
\node at (1.7,-0.3) {\( 0,j \)};
\node at (1.0,-0.3) {\( t,i \)};
\node at (0.,-0.3) {\( 0,j \)};
\node at (1.3,0.) {\( \simeq \)};
\begin{feynman}
    \vertex (a1);
    \vertex [right=1.0cm of a1] (a2);
    \vertex [right=0.7cm of a2] (a3);
    \vertex [right=1.0cm of a3] (a4){~+~};
    \vertex [right=0.4cm of a4] (a5);
    \vertex [right=0.7cm of a5] (a6);
    \vertex [right=1.0cm of a6] (a7);
    \vertex [right=0.7cm of a7] (a8){~+~};
    \vertex [right=0.4cm of a8] (a9);
    \vertex [right=0.7cm of a9] (a10);
    \vertex [right=0.7cm of a10] (a11);
    \vertex [right=0.7cm of a11] (a12);
    \vertex [right=0.7cm of a12] (a13);
    \vertex [right=0.7cm of a13] (a14){,};
    \diagram*{
      (a1) --[double] (a2),  
      (a3) --[plain] (a4),  
      (a5) --[plain] (a6),
      (a6) --[double] (a8),      
      (a6) -- [scalar, out=90, in=90, looseness=2.0,thick] (a7),
      (a9) --[plain] (a10),      
      (a10) --[double] (a14),    
      (a10) -- [scalar, out=90, in=90, looseness=1.5,thick] (a12),
      (a11) -- [scalar, out=90, in=90, looseness=1.5,thick] (a13),
    };
  \end{feynman}
\end{tikzpicture}
\end{equation}
which can be approximated by substituting internal double lines for zigzags:
\begin{equation}\label{eqn:iter-bis}
\begin{tikzpicture}
\node at (10.0,-0.3) {\( t,i \)};
\node at (9.3,-0.3) {\( t_1 \)};
\node at (8.6,-0.3) {\( t_2 \)};
\node at (7.9,-0.3) {\( t_3 \)};
\node at (7.2,-0.3) {\( t_4 \)};
\node at (6.5,-0.3) {\( 0,j \)};
\node at (5.8,-0.3) {\( t,i \)};
\node at (5.1,-0.3) {\( t_1 \)};
\node at (4.1,-0.3) {\( t_2 \)};
\node at (3.4,-0.3) {\( 0,j \)};
\node at (2.7,-0.3) {\( t,i \)};
\node at (1.7,-0.3) {\( 0,j \)};
\node at (1.0,-0.3) {\( t,i \)};
\node at (0.,-0.3) {\( 0,j \)};
\node at (1.3,0.) {\( \simeq \)};
\begin{feynman}
    \vertex (a1);
    \vertex [right=1.0cm of a1] (a2);
    \vertex [right=0.7cm of a2] (a3);
    \vertex [right=1.0cm of a3] (a4){~+~};
    \vertex [right=0.4cm of a4] (a5);
    \vertex [right=0.7cm of a5] (a6);
    \vertex [right=1.0cm of a6] (a7);
    \vertex [right=0.7cm of a7] (a8){~+~};
    \vertex [right=0.4cm of a8] (a9);
    \vertex [right=0.7cm of a9] (a10);
    \vertex [right=0.7cm of a10] (a11);
    \vertex [right=0.7cm of a11] (a12);
    \vertex [right=0.7cm of a12] (a13);
    \vertex [right=0.7cm of a13] (a14){.};
    \diagram*{
      (a1) --[double] (a2),  
      (a3) --[plain] (a4),  
      (a5) --[plain] (a6),
      (a6) --[boson] (a7),      
      (a7) --[double] (a8),      
      (a6) -- [scalar, out=90, in=90, looseness=2.0,thick] (a7),
      (a9) --[plain] (a10),      
      (a10) --[boson] (a11),    
      (a11) --[boson] (a12),    
      (a12) --[boson] (a13),    
      (a13) --[double] (a14),    
      (a10) -- [scalar, out=90, in=90, looseness=1.5,thick] (a12),
      (a11) -- [scalar, out=90, in=90, looseness=1.5,thick] (a13),
    };
  \end{feynman}
\end{tikzpicture}
\end{equation}
In this manner, both the nested and crossed contributions give rise to linear terms in $\mathcal{L}[\langle w_i(x)\rangle](s)$:
\begin{equation}\label{eqn:iter-n-laplace}
\begin{tikzpicture}
\node at (2.4,-0.3) {\( i \)};
\node at (1.7,-0.3) {\( t_1 \)};
\node at (0.8,-0.3) {\( t_2 \)};
\node at (0.,-0.3) {\( j \)};
\node at (9.2,0.05) {\( =~\displaystyle \frac{\delta\Omega^2}{3}\mathcal{L}[\langle w_i(x)\rangle](s)\mathcal{L}[1](s)
\mathcal{L}\left[\left[R^{-1}_{ij}(2\Omega_0x) -(1+2\cos\Omega_0x)R^{-1}_{ij}(\Omega_0x)\right]\varphi(x)\langle w_i(x)\rangle_0\right](s), \)};
\begin{feynman}
    \vertex (a1);
    \vertex [right=0.7cm of a1] (a2);
    \vertex [right=1cm of a2] (a3);
    \vertex [right=0.7cm of a3] (a4);
    \diagram*{
      (a1) --[plain] (a2),  
      (a2) --[boson] (a3),  
      (a3) --[double] (a4),  
      (a2) -- [scalar, out=90, in=90, looseness=2.0,thick] (a3),
    };
  \end{feynman}
\end{tikzpicture}
\end{equation}
\begin{equation}\label{eqn:iter-c-laplace}
\begin{tikzpicture}
\node at (3.6,-0.3) {\( i \)};
\node at (2.9,-0.3) {\( t_1 \)};
\node at (2.2,-0.3) {\( t_2 \)};
\node at (1.5,-0.3) {\( t_3 \)};
\node at (0.8,-0.3) {\( t_4 \)};
\node at (0.,-0.3) {\( j \)};
\node at (9.7,0) {\( ~~~=~ \displaystyle \left(\frac{\delta\Omega^2}{3}\right)^2\mathcal{L}[\langle w_i(x)\rangle](s)\mathcal{L}[1](s)
\mathcal{L}\bigg[\big[1+R^{-1}_{ij}(\Omega_0(x_1+2x_2+x_3))(1+2\cos\Omega_0(x_1-x_3)) \)};
\node at (9.0,-1) {\(  - R^{-1}_{ij}(\Omega_0(2x_1+2x_2)) - R^{-1}_{ij}(\Omega_0(2x_2+2x_3))\big]\varphi(x_1+x_2)\varphi(x_2+x_3)\langle w_i(x_1)\rangle_0\langle w_i(x_2)\rangle_0\langle w_i(x_3)\rangle_0\bigg](s). \)};
\begin{feynman}
    \vertex (a1);
    \vertex [right=0.7cm of a1] (a2);
    \vertex [right=0.7cm of a2] (a3);
    \vertex [right=0.7cm of a3] (a4);  \vertex [right=0.7cm of a4] (a5);
    \vertex [right=0.7cm of a5] (a6);{~=~};
    \diagram*{
      (a1) --[plain] (a2),  
      (a2) --[boson] (a5),  
      (a5) --[double] (a6),  
      (a2) -- [scalar, out=90, in=90, looseness=1.5,thick] (a4),
      (a3) -- [scalar, out=90, in=90, looseness=1.5,thick] (a5),
    };
  \end{feynman}
\end{tikzpicture}
\end{equation}
Note that the solution obtained for $\langle w_i(t)\rangle$ can be used as a starting solution for iterating further based on Eqn.~\ref{eqn:iter-bis}. 
\end{widetext}

\section{Parallel diffusion}
\label{sec:parallel}

The velocity decorrelation function relevant for the parallel diffusion corresponds to $i=j=z$ in equations of Section~\ref{sec:summation}. Denoting for convenience as $\hat{W}_{i}(s)$ the Laplace transform function $\mathcal{L}[\langle w_{i}(t)\rangle](s)$, the first iterated propagator is inferred from Eqn.~\ref{eqn:bourret} in the Laplace space that reads as
\begin{equation}
\label{eqn:W0-zz}
    \hat{W}_{0z}(s)=\frac{1}{s}-\frac{2\delta\Omega^2}{3}\frac{\hat{W}_{0z}(s)}{s}\mathcal{L}\left[\varphi(x)\cos\Omega_0x\right](s),
\end{equation}
while the iterated propagator is inferred from Eqn.~\ref{eqn:iter-bis} as
\begin{multline}
\label{eqn:W-zz}
    \hat{W}_z(s)=\frac{1}{s}-\frac{2\delta\Omega^2}{3}\frac{\hat{W}_z(s)}{s}\mathcal{L}\left[\varphi(x)\langle w_z(x)\rangle_0\cos\Omega_0x\right](s)\\
    +2\left(\frac{\delta\Omega^2}{3}\right)^2\frac{\hat{W}_z(s)}{s}\mathcal{L}\bigg[\varphi(x_1+x2)\varphi(x_2+x3)\\
    \langle w_z(x_1)\rangle_0\langle w_z(x_2)\rangle_0\langle w_z(x_3)\rangle_0\cos\Omega_0(x_1-x_3)\bigg](s).
\end{multline}

\begin{figure}[t]
\centering
\includegraphics[width=\columnwidth]{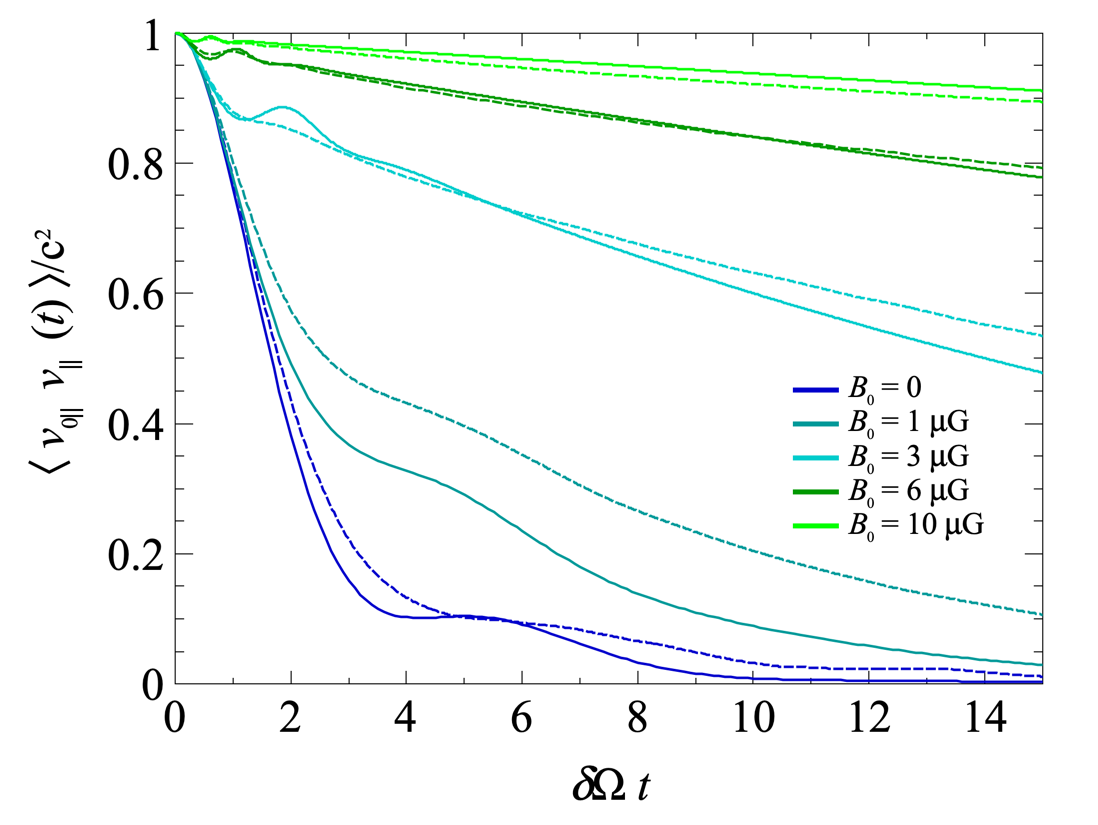}
\caption{Parallel velocity decorrelation function of particles with reduced rigidity $\rho=0.1$ for different values of $B_0$ ($\delta B=1~\upmu$G), as a function of the gyroperiod time scale $\delta\Omega t$. Dashed lines are from Monte-Carlo simulations.}
\label{fig:vv-par-rho0.1}
\end{figure}

The resulting velocity decorrelation functions $\langle v_{0\parallel}v_\parallel(t)\rangle=\langle w_{0z}w_z(t)\rangle$ are shown in Fig.~\ref{fig:vv-par-rho0.1} for $\rho=0.1$ and different values of $B_0$ (and $\delta B=1~\upmu$G). For reference, results from Monte-Carlo experiments are shown as the dashed lines. Overall, the main features of the functions inferred from the simulations, namely the modulations on top of an approximately exponential envelope that is decreasing slower with time for increasing $B_0$ values, are captured by the calculation. In this rigidity regime, the resonance between the Larmor radius of the particles with wavelengths of the turbulence is the source of the modulations related to the total angular frequency $\delta\Omega+\Omega_0$. They reflect memory effects originating from large-scale wavenumber field lines around which particles spiral while undergoing the imprint of a random walk caused by smaller wavenumber vectors. As the intensity $B_0$ increases, the particles tend to remain bound to the lines of the mean field for longer, and the decay takes longer. Beyond similarities between the simulation and calculation results, however, quantitative differences are observed in Fig.~\ref{fig:vv-par-rho0.1}. The most notable one concerns the global rate of falloff of the decorrelation functions that is predicted to be too rapid for $B_0\lesssim 5\upmu$G by the calculation compared to the Monte-Carlo simulations. The increase of the ``decay time'' describing the approximately exponential envelope is indeed too slow for small values of $B_0$, as clearly observed for $B_0=1\upmu$G, before to cross the right range and to get too fast for $B_0\gtrsim 5\upmu$G. In other words, the dependence of the decorrelation functions in $B_0$ is non-linearly under-(over)estimated for $B_0\lesssim (\gtrsim) 5\upmu$G. The underestimation for small $B_0$, already visible for $B_0=0$, is attributed at this stage on the one hand to the overestimation of $\varphi(t)$ observed in Fig.~\ref{fig:dbdb} for $\rho=0.1$, and on the other hand to an artefact due to the partial summation of the Dyson series. 

\begin{figure}[t]
\centering
\includegraphics[width=\columnwidth]{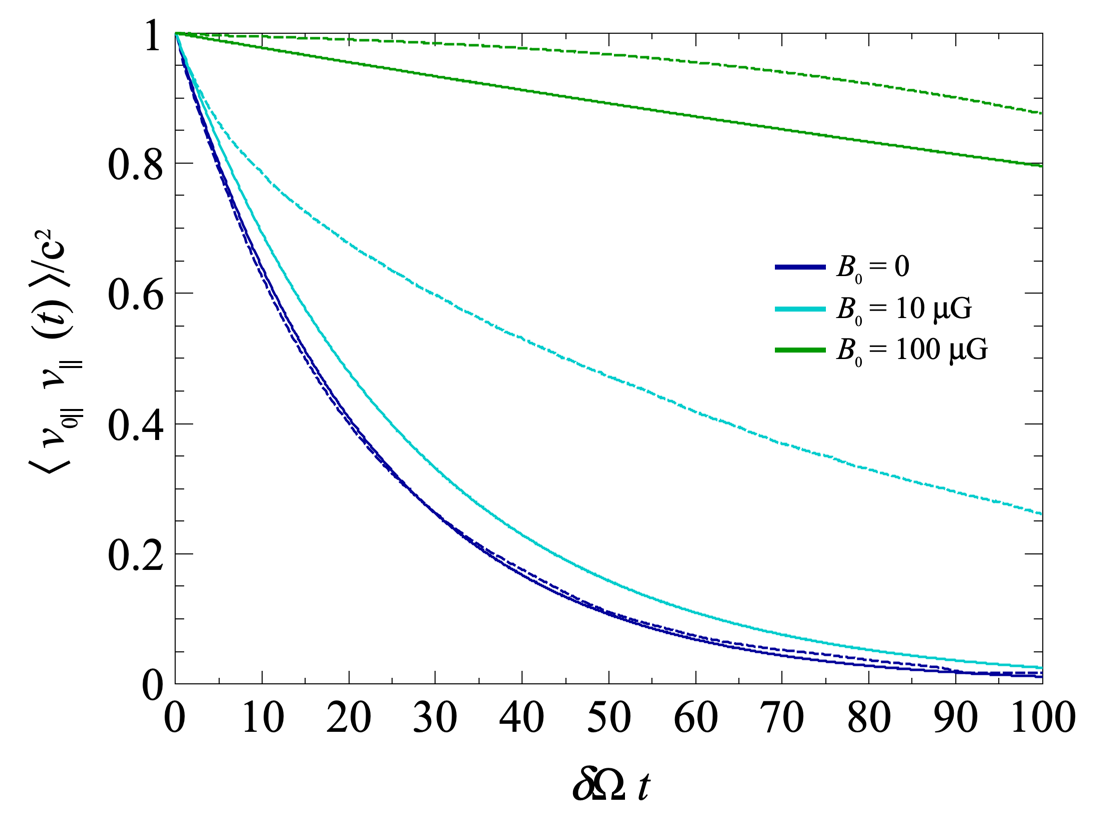}
\caption{Same as Fig.~\ref{fig:vv-par-rho0.1} for $\rho=1$.}
\label{fig:vv-par-rho-1}
\end{figure}

At higher rigidity ($\rho=1$), the Larmor radius of the particles is always larger than the eddy sizes and scatterings can be considered independent one from another. The process is Markovian and the decorrelation gets exponential in the pure turbulence case, as demonstrated in~\cite{Plotnikov:2011me} based on a white-noise approximation to describe the 2-pt correlation function $\varphi(t)$. By increasing $B_0$, the decay of the decorrelations gets slower. Similarly to the case $\rho=0.1$, the calculation is observed to underestimate the ``decay time''; yet the values of $B_0$ leading to the underestimation span a much wider range. Because the overestimation of $\varphi(t)$ observed in Fig.~\ref{fig:dbdb} for $\rho=1$ is rather small, the velocity decorrelation function coincides quite well with the Monte-Carlo one for $B_0=0$. Consequently, the differences for $B_0>0$ stem predominantly from some incompleteness in the partial summation of the Dyson series.

\begin{figure}[t]
\centering
\includegraphics[width=\columnwidth]{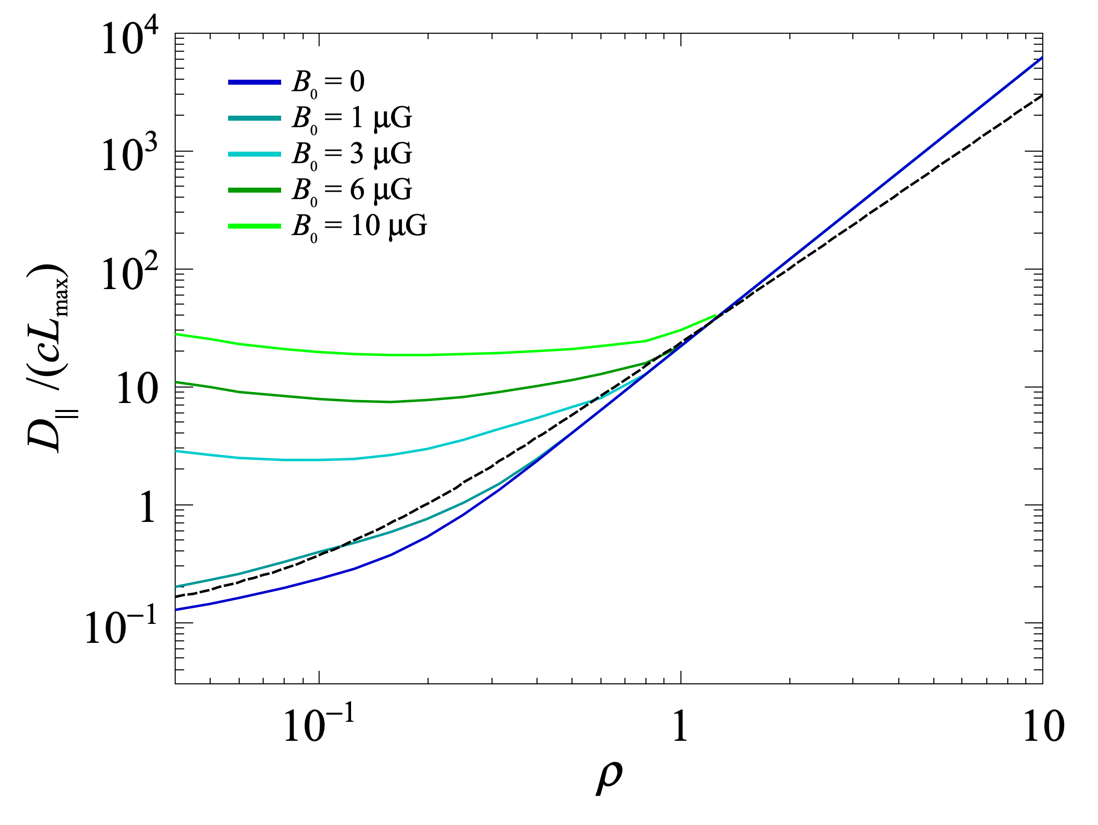}
\caption{Parallel diffusion coefficient as a function of the reduced rigidity for different values of $B_0$ ($\delta B=1~\upmu$G). The Dashed line is from Monte-Carlo simulations in the pure turbulent case ($B_0=0)$.}
\label{fig:D-par}
\end{figure}

The dependence in rigidity of the parallel diffusion coefficient $D_\parallel$ as obtained from Eqn.~\ref{eqn:Dij} is shown in Fig.~\ref{fig:D-par} for different values of $B_0$, expressed in units of $cL_{\mathrm{max}}$. The dashed line displays, for reference, the results obtained from Monte-Carlo simulations in the case of pure turbulence. Despite the aforementioned differences between the simulations and the calculation, $D_\parallel$ is observed to be reproduced within a factor 2. In particular, the calculation slightly deviates from the expected scalings in $\rho^{1/3}$ in the gyroresonant regime an in $\rho^2$ in the quasi-ballistic regime~\cite{Wentzel:1974cp,Bhattacharjee:1999mup,Aloisio:2004jda}. As $B_0$ is increasing, from the analysis of the decorrelation functions presented above, $D_\parallel$ is expected to be more and more underestimated; yet the calculation presented here provides genuine qualitative scalings that are also reliable quantitatively around $\rho=0.1$.

\section{Perpendicular diffusion}
\label{sec:perpendicular}

The velocity decorrelation function relevant for the perpendicular diffusion corresponds to $i=j=x$ or $i=j=y$ in equations of Section~\ref{sec:summation}. The calculation proceeds the same way as for the parallel diffusion:
\begin{equation}
\label{eqn:W0-xx}
    \hat{W}_{0x}(s)=\frac{1}{s}-\frac{\delta\Omega^2}{3}\frac{\hat{W}_{0x}(s)}{s}\mathcal{L}\left[\varphi(x)\left(1+\cos{\Omega_0x}\right)\right](s),
\end{equation}
and
\begin{multline}
\label{eqn:W-xx}
    \hat{W}_x(s)=\frac{1}{s}-\frac{\delta\Omega^2}{3}\frac{\hat{W}_x(s)}{s}\mathcal{L}\left[\varphi(x)\langle w_x(x)\rangle_0(1+\cos\Omega_0x)\right](s)\\
    +2\left(\frac{\delta\Omega^2}{3}\right)^2\frac{\hat{W}_x(s)}{s}\mathcal{L}\bigg[
    \varphi(x_1+x2)\varphi(x_2+x3)\\
    \langle w_x(x_1)\rangle_0\langle w_x(x_2)\rangle_0\langle w_x(x_3)\rangle_0\cos^2{\Omega_0\left(\frac{x_1}{2}+x_2+\frac{x_3}{2}\right)}\bigg](s).
\end{multline}
The relevant decorrelation function is $\langle v_{0\perp}v_\perp(t)\rangle=\langle w_{0x}w_x(t)\rangle\cos\Omega_0t=\langle w_{0y}w_y(t)\rangle\cos\Omega_0t$. 

\begin{figure}[h]
\centering
\includegraphics[width=\columnwidth]{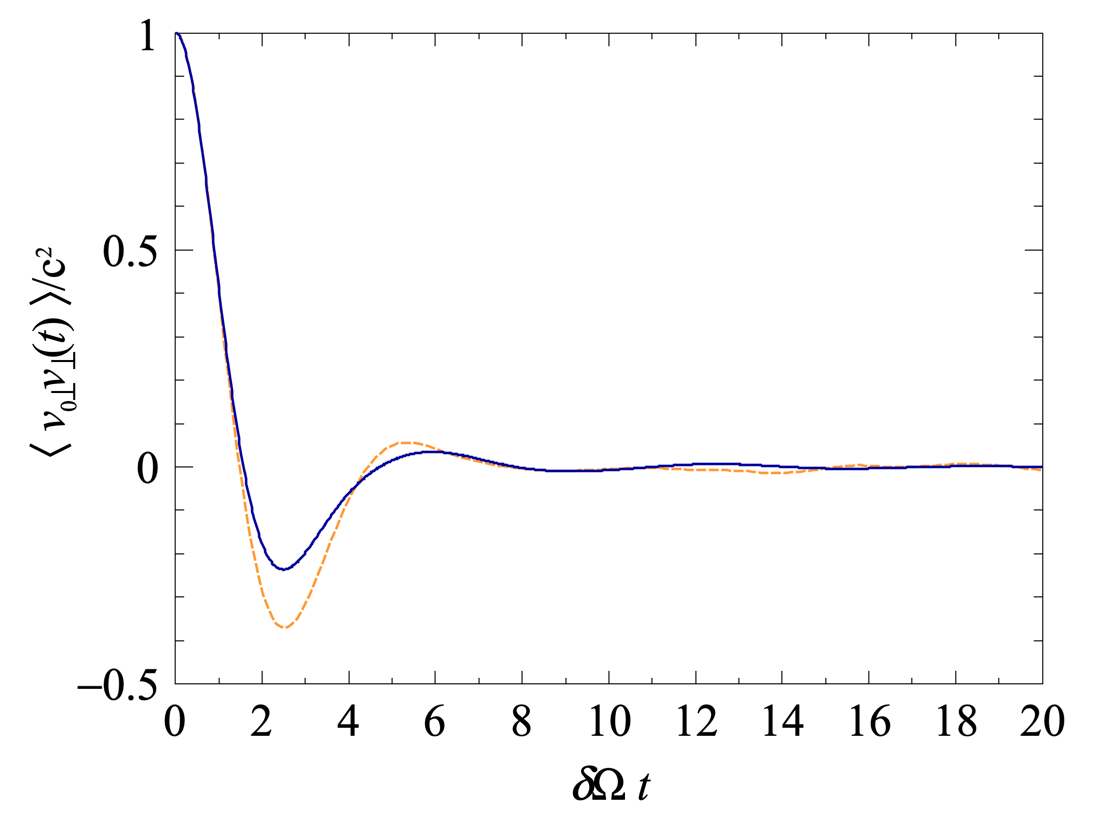}
\caption{Perpendicular velocity decorrelation function of particles with reduced rigidity $\rho=0.1$ for $\delta B=1~\upmu$G and $B_0=1\upmu$G, as a function of the gyroperiod time scale $\delta\Omega t$. Dashed line is from Monte-Carlo simulations.}
\label{fig:vv-perp-rho-0.1-1}
\end{figure}

A first illustration of the calculation is given in Fig.~\ref{fig:vv-perp-rho-0.1-1}, where the perpendicular decorrelation function is shown for $\rho=0.1$ and $B_0=1~\upmu$G. The dashed line is from Monte-Carlo simulations. As in the case of parallel diffusion in Section~\ref{sec:parallel}, the envelope of the decay is slightly underestimated by the calculation. However, the modulations features, which have been shown from various simulations to be responsible for the perpendicular sub-diffusive regime at early times, are well reproduced. It is to be noted that none of the approximations proposed in the literature could predict both the fast decaying envelope and the positions of minimum and second maximum inherited from the modulations on top of the decay. 

\begin{figure}[h]
\centering
\includegraphics[width=\columnwidth]{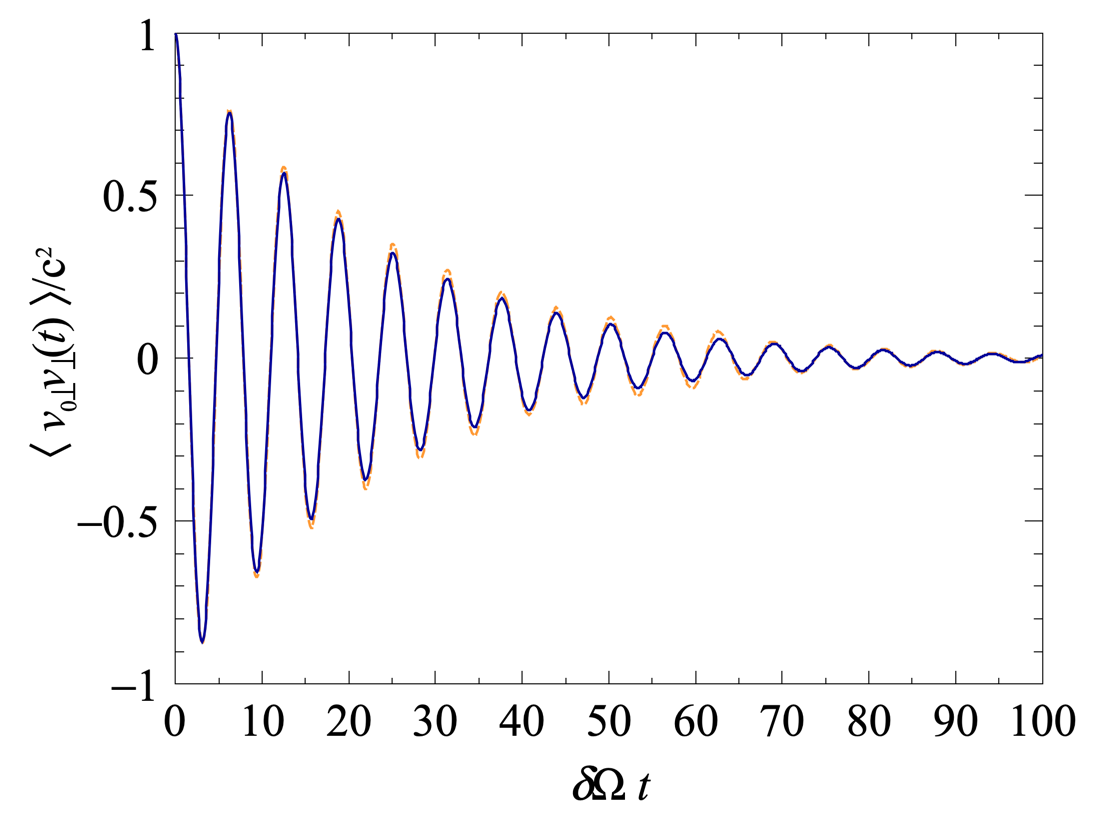}
\caption{Same as Fig.~\ref{fig:vv-perp-rho-0.1-1} for $\rho=1$ and $B_0=3\upmu$G.}
\label{fig:vv-perp-rho-1-1}
\end{figure}

\begin{figure}[t]
\centering
\includegraphics[width=\columnwidth]{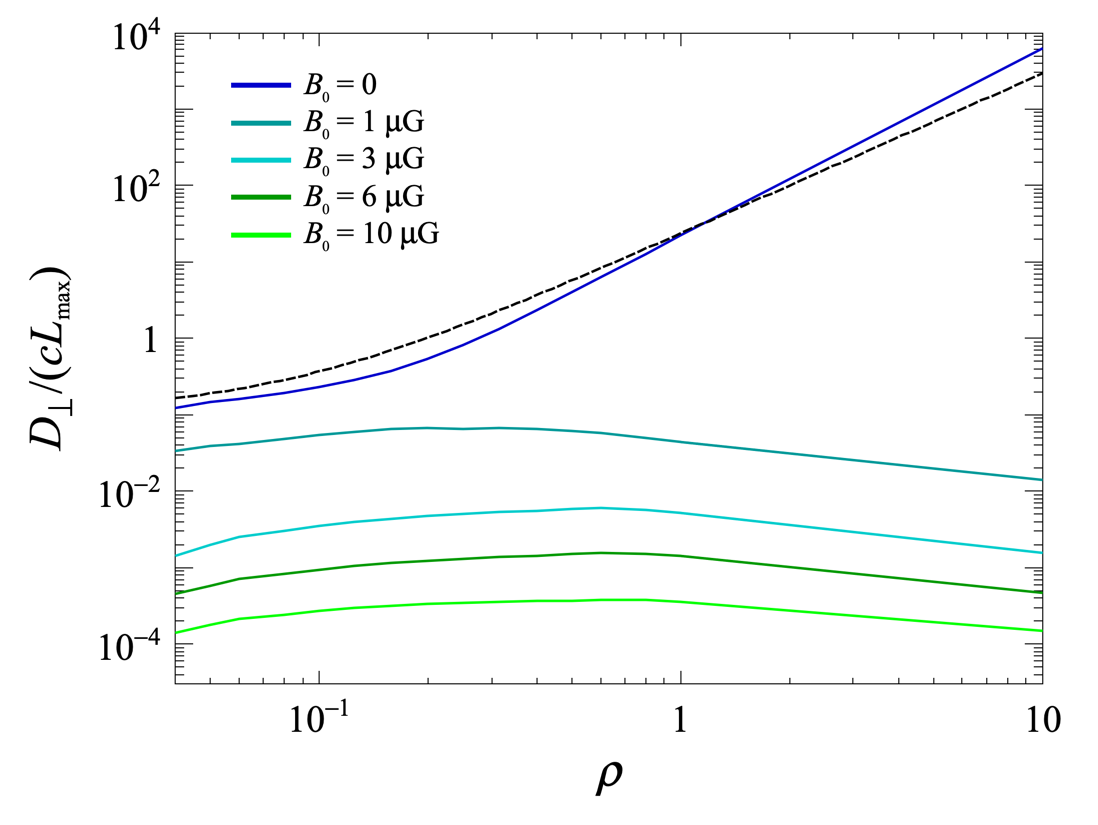}
\caption{Perpendicular diffusion coefficient as a function of the reduced rigidity for different values of $B_0$ ($\delta B=1~\upmu$G). The Dashed line is from Monte-Carlo simulations in the pure turbulent case ($B_0=0)$.}
\label{fig:D-perp}
\end{figure}

\begin{figure}[t]
\centering
\includegraphics[width=\columnwidth]{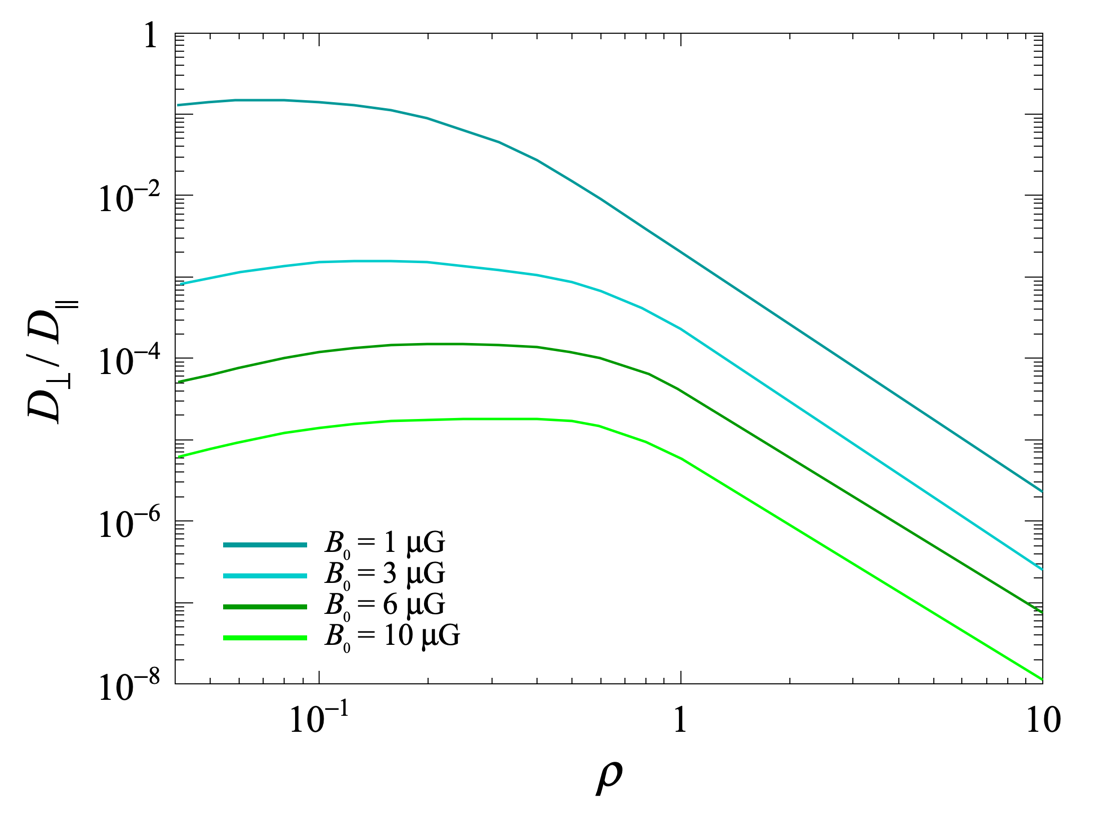}
\caption{Ratio between perpendicular and parallel diffusion coefficient as a function of the reduced rigidity for different values of $B_0$ ($\delta B=1~\upmu$G).}
\label{fig:ratio-perp-par}
\end{figure}

A second illustration is given in Fig.~\ref{fig:vv-perp-rho-1-1}, for $\rho=0.1$ and $B_0=1~\upmu$G. At such a high rigidity, the decay time scale is longer than the gyroperiod time scale $\Omega_0t$, hence the numerous oscillations. The results from the calculation or from the simulations are almost indistinguishable. 

From these two illustrations, we observe that the successes based on Eqn.~\ref{eqn:W-xx} follow those presented in Section~\ref{sec:parallel} in the case of parallel diffusion: the calculation is able to reproduce the main features uncovered by numerical simulations.  However, the same limitations apply to the scaling of the results with $B_0$, as a function of $\rho$.  

In the same manner as in the parallel transport, the dependence in rigidity of the perpendicular diffusion coefficient $D_\perp$ is shown in Fig.~\ref{fig:D-perp} for different values of $B_0$, expressed in units of $cL_{\mathrm{max}}$. Due to the oscillatory behavior of the velocity decorrelation functions, the additional time integration should smooth out differences between the simulations and the calculations. As $B_0$ is increasing, $D_\perp$ is decreasing, as expected (for $B_0\rightarrow\infty$, a  particle would be spiraling around $\mathbf{B}_0$ at a fixed radius). More revealing is the rigidity dependence of $D_\perp$, observed to rise more slowly than that of $D_\parallel$ in the gyroresonant regime and, unlike $D_\parallel$, to decrease in the high-rigidity regime. These dependencies are more clearly highlighted in Fig.~\ref{fig:ratio-perp-par}, where the ratio $D_\perp/D_\parallel$ is displayed. The rise at low rigidities is in agreement with that revealed in Monte-Carlo studies in which the turbulence dynamical range well covers the rigidities of interest~\cite{Dundovic:2020sim}. Furthermore, the decrease at high rigidities is also in agreement with these simulations, as is the shift of the transition region towards higher rigidities as $B_0$ is increasing. 

\section{Anti-symmetric diffusion}
\label{sec:anti-symmetric}

\begin{figure}[h]
\centering
\includegraphics[width=\columnwidth]{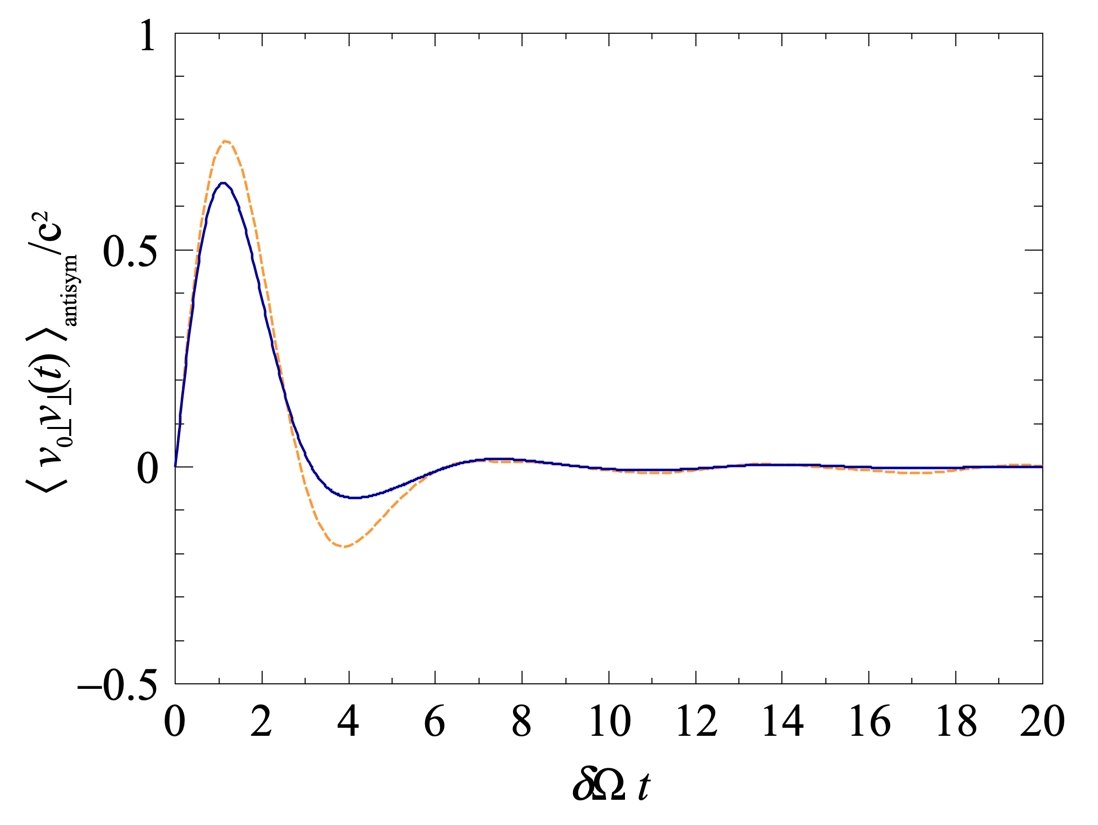}
\caption{Anti-symmetric velocity decorrelation function of particles with reduced rigidity $\rho=0.1$ for $\delta B=1~\upmu$G and $B_0=1\upmu$G, as a function of the gyroperiod time scale $\delta\Omega t$. Dashed line is from Monte-Carlo simulations.}
\label{fig:vv-hall-rho-0.1-1}
\end{figure}

Finally, we provide the velocity decorrelation function relevant for the anti-symmetric diffusion by illustrating the case $j=y,i=x$:
\begin{equation}
\label{eqn:W0-xx-anti}
    \hat{W}_{0x}(s)=\frac{1}{s}-\frac{\delta\Omega^2}{3}\frac{\hat{W}_{0x}(s)}{s}\mathcal{L}\left[\varphi(x)\sin{\Omega_0x}\right](s),
\end{equation}
and
\begin{multline}
\label{eqn:W-xx-anti}
    \hat{W}_x(s)=\frac{1}{s}-\frac{\delta\Omega^2}{3}\frac{\hat{W}_x(s)}{s}\mathcal{L}\left[\varphi(x)\langle w_x(x)\rangle_0\sin{\Omega_0x}\right](s)\\
    +2\left(\frac{\delta\Omega^2}{3}\right)^2\frac{\hat{W}_x(s)}{s}\mathcal{L}\bigg[
    \varphi(x_1+x2)\varphi(x_2+x3)\langle w_x(x_1)\rangle_0\\
    \langle w_x(x_2)\rangle_0\langle w_x(x_3)\rangle_0\left(1-\sin{\Omega_0\left(x_1+2x_2+x_3\right)}\right)\bigg](s).
\end{multline}
The relevant decorrelation function is $\langle v_{0\perp}v_\perp(t)\rangle_{\mathrm{antisym}}=\langle w_{0x}w_y(t)\rangle\sin\Omega_0t=-\langle w_{0y}w_x(t)\rangle\sin\Omega_0t$. Similarly with the perpendicular velocity decorrelation case of Section~\ref{sec:perpendicular}, the  envelope of the decay is slightly underestimated by the calculation compared to the Monte-Carlo results in the case $\rho=0.1$ and $B_0=1~\upmu$G. 

Finally, the dependence in rigidity of the anti-symmetric diffusion coefficient $D_{\mathrm{a.-s.}}$ is observed to be linear, while that in $B_0$ is observed to be inversely proportional. Overall, the shorthand expression
\begin{equation}
    \frac{D_{\mathrm{a.-s.}}}{cL_\mathrm{max}\rho}\simeq \frac{\delta B}{B_0}
\end{equation}
is enough to capture the dependencies in the range of $B_0$ studied, $0.1\leq \delta B/B_0\leq 1$.

\section{Conclusion} 
\label{sec:discussion}

Velocity decorrelation functions of high-energy cosmic rays propagating in magnetic fields have been obtained from the Dyson series governing the motion of the particles. The absence of small parameter of expansion forbids any perturbation theory to hold, and hence any truncation of the Dyson series~\cite{Kraichnan:1961:DNS}. The partial-summation scheme developed in this study is shown to provide an approximate solution that captures the main features uncovered by numerical simulations in a range of rigidities covering the transition between the gyroresonant and the quasi-ballistic regimes on the one hand, and the quasi-ballistic regime itself on the other hand. Keeping in mind the limitations underlined in terms of underestimation of the diffusion coefficients, the calculation based on Eqn.~\ref{eqn:dyson-bis} approximated by Eqn.~\ref{eqn:iter-bis} provides a rapid tool for deriving approximate solutions without having to resort to heavy numerical simulation campaigns.  

The calculation was illustrated in the case of 3D isotropic turbulence following a Kolmogorov power spectrum. This turbulence model is particularly useful as a reference, as it has been widely used in the literature as a case study for Monte Carlo simulations of test particles. However, the computational techniques presented in this article are not limited to this particular turbulence. Several astrophysical contexts require a 1D ``slab'' turbulence model, a 2D isotropic one, 3D anisotropic ones with or without helicity~\cite{1995ApJ...438..763G,PhysRevLett.88.245001,2003MNRAS.345..325C}. All these turbulence are described by different spectral tensors $P_{ij}$ compared to the one used in Eqn.~\ref{eqn:Pij} that would give specific expressions for the 2-pt function of the magnetic field compared to Eqn.~\ref{eqn:dbdb-iso3D}. A comprehensive study of these turbulence models is planned in a future study.

A key ingredient of the approximate solution relies on the modeling of the 2-pt correlation function of the turbulence experienced by the particles between two successive times. The approximation proposed in this study is found to reproduce the main features uncovered, here again, by numerical simulations as long as the reduced rigidity is larger than about one tenth of the largest scale of the turbulence. Progress in the modeling of this 2-pt function is needed to extend to lower rigidity the range of validity of the type of calculation presented in this study.

\acknowledgments

I gratefully thank Patrick Reichherzer and Leander Schlegel for providing me the Monte-Carlo simulations presented in Fig.~\ref{fig:dbdb}, and acknowledge funding from ANR via the grant MultI-messenger probe of Cosmic Ray Origins (MICRO), ANR-20-CE92-0052.  This work was also made possible with the support of the Institut Pascal at Universit\'e Paris-Saclay during the Paris-Saclay Astroparticle Symposium 2022, with the support of the P2IO Laboratory of Excellence (program ``Investissements d’avenir'' ANR-11-IDEX-0003-01 Paris-Saclay and ANR-10-LABX-0038), the P2I axis of the Graduate School Physics of Universit\'e Paris-Saclay, as well as IJCLab, CEA, IPhT, APPEC, the IN2P3 master projet UCMN and EuCAPT ANR-11-IDEX-0003-01 Paris-Saclay and ANR-10-LABX-0038). Finally, I thank Haris Lyberis for his help on an earlier development of this work.

\appendix

\section{Monte-Carlo generator}
\label{app:MC}

To serve as a reference for testing the model, 
a Monte-Carlo estimation of the velocity correlation function is used. Following a scheme similar to that widely used in the literature, a large number of particle trajectories in given  magnetic field configurations is simulated by solving numerically Eqn.~\ref{eqn:LorentzNewton}. The numerical integration is performed using the standard Runge-Kutta integrator. 

The regular field is oriented along the $z$ axis, $\mathbf{B}_0=B_0\mathbf{u}_z$, with $B_0$ constant in a given configuration. On top of this constant field, a turbulence $\delta\mathbf{B}$ is added. To simulate numerically an isotropic and spatially homogeneous turbulent field, an algorithm similar to that in~\cite{Batchelor70,1999ApJ...520..204G} is used. The recipe consists in summing a large number $N_\mathrm{m}$ of plane waves ($N_\mathrm{m}=250$ in this study) with corresponding wave vector $\mathbf{k}_n$, the direction, phase $\phi_n$ and polarisation of which are chosen randomly:
\begin{equation}
\label{eqn:dBMC}
\delta\mathbf{B}(\mathbf{x})=\sqrt{2}\sum_{n=1}^{N_\mathrm{m}} \sum_{\alpha=1}^{2} \mathcal{E}_n(k_n)~\mathbf{\hat{\xi}}_n^\alpha~\cos{(\mathbf{k}_n\cdot\mathbf{x}+\phi_n^\alpha)}.
\end{equation}
To ensure the condition $\nabla\cdot\delta \mathbf{B}=0$, the two orthogonal polarisation vectors $\mathbf{\hat{\xi}}_n^\alpha$ are oriented in the plane perpendicular to the directions of the wave vectors. The wave number distribution is built from a constant logarithmic spacing between $k_{\mathrm{min}}$ and $k_{\mathrm{max}}$. The wave amplitudes satisfy \mbox{$\mathcal{E}_n^2(k_n)=\mathcal{E}_0\delta B^2k_n^{-5/3}(k_n-k_{n-1})$}, where $\mathcal{E}_0$ is a normalisation factor such that $\sum_n \mathcal{E}_n^2(k_n)=\delta B^2$. In this manner, the turbulence satisfies $\langle\delta\mathbf{B}(\mathbf{x})\rangle=0$ and \mbox{$\langle \delta\mathbf{B}_i(\mathbf{x})\delta\mathbf{B}_j(\mathbf{x}')\rangle=\delta B^2\delta_{ij}$}. The dynamic range of the turbulence explored here is \mbox{$L_\mathrm{max}/L_\mathrm{min}=100$}.

\section{Perpendicular and anti-symmetric contributions for $p=2$}
\label{app:p=2}

For completeness and clarity about the calculation rules of the diagrams, explicit expressions of the perpendicular and anti-symmetric contributions are provided in this Appendix in the case $p=2$. 

\begin{widetext}
Starting from the general expression of the crossed diagrams,
\begin{equation}\label{eqn:diagram-4c}
\begin{tikzpicture}
\node at (3.1,-0.3) {\( t,i_0 \)};
\node at (2.6,-0.3) {\( t_1 \)};
\node at (1.9,-0.3) {\( t_2 \)};
\node at (1.2,-0.3) {\( t_3 \)};
\node at (0.6,-0.3) {\( t_4 \)};
\node at (0.,-0.3) {\( 0,i_4 \)};
\node at (7.8,0.05) {\( =\displaystyle ~\left(\frac{\delta\Omega^2}{3}\right)^2\displaystyle\int_0^t\dif t_1\int_0^{t_1}\dif t_2\int_0^{t_2}\dif t_3\int_0^{t_3}\dif t_4~\varphi(t_1-t_3)\varphi(t_2-t_4) \)};
\node at (7.8,-0.95) {\( \times\displaystyle ~\left( R_{i_0k_1}(\Omega_0t_1) R_{i_2k_1}(\Omega_0t_3)R^{-1}_{m_1i_1}(\Omega_0t_1)R^{-1}_{m_1i_3}(\Omega_0t_3)- R_{i_0k_1}(\Omega_0t_1) R_{i_2k_3}(\Omega_0t_3)R^{-1}_{k_3i_1}(\Omega_0t_1)R^{-1}_{k_1i_3}(\Omega_0t_3)\right) \)};
\node at (7.9,-1.95) {\( \times\displaystyle ~\left( R_{i_1k_2}(\Omega_0t_1) R_{i_3k_2}(\Omega_0t_3)R^{-1}_{m_2i_2}(\Omega_0t_1)R^{-1}_{m_2i_4}(\Omega_0t_3)- R_{i_1k_2}(\Omega_0t_1) R_{i_3k_4}(\Omega_0t_3)R^{-1}_{k_4i_2}(\Omega_0t_1)R^{-1}_{k_2i_4}(\Omega_0t_3)\right), \)};
\begin{feynman}
    \vertex (a1);
    \vertex [right=0.5cm of a1] (a2);
    \vertex [right=0.7cm of a2] (a3);
    \vertex [right=0.7cm of a3] (a4);
    \vertex [right=0.7cm of a4] (a5);
    \vertex [right=0.5cm of a5] (a6);
    \diagram*{
      (a1) --[plain] (a6),  
      (a2) -- [scalar, out=90, in=90, looseness=1.5,thick] (a4),
      (a3) -- [scalar, out=90, in=90, looseness=1.5,thick] (a5),
    };
\end{feynman}
\end{tikzpicture}
\end{equation}
and using rotation-matrix properties such as $R_{ij}(x_1)R^{-1}_{jk}(x_2)=R_{ik}(x_1-x_2)$, $R_{ij}(x)=R_{ji}(-x)=R^{-1}_{ij}(-x)$ and $R_{ii}(x)=R^{-1}_{ii}(x)=1+2\cos\Omega_0x$, we obtain the expressions of the perpendicular and anti-symmetric diagrams:
\begin{equation}\label{eqn:diagram-4c-xx}
\begin{tikzpicture}
\node at (7.2,-0.3) {\( t,y \)};
\node at (6.6,-0.3) {\( t_1 \)};
\node at (6.0,-0.3) {\( t_2 \)};
\node at (5.3,-0.3) {\( t_3 \)};
\node at (4.7,-0.3) {\( t_4 \)};
\node at (4.1,-0.3) {\( 0,y \)};
\node at (3.6,-0.) {\( = \)};
\node at (3.1,-0.3) {\( t,x \)};
\node at (2.6,-0.3) {\( t_1 \)};
\node at (1.9,-0.3) {\( t_2 \)};
\node at (1.2,-0.3) {\( t_3 \)};
\node at (0.6,-0.3) {\( t_4 \)};
\node at (0.,-0.3) {\( 0,x \)};
\node at (7.85,-1.05) {\( =\displaystyle ~\left(\frac{\delta\Omega^2}{3}\right)^2\displaystyle\int_0^t\dif t_1\int_0^{t_1}\dif t_2\int_0^{t_2}\dif t_3\int_0^{t_3}\dif t_4~\varphi(t_1-t_3)\varphi(t_2-t_4) \)};
\node at (7.8,-2.05) {\( \times\displaystyle ~\bigg( 1-\cos{2\Omega_0(t_1-t_3)}-\cos{2\Omega_0(t_2-t_4)} + \cos{\Omega_0(t_1+t_2-(t_3+t_4))}[1+2\cos{\Omega_0(t_2+t_3-(t_1+t_4))}] \bigg), \)};
\begin{feynman}
    \vertex (a1);
    \vertex [right=0.5cm of a1] (a2);
    \vertex [right=0.7cm of a2] (a3);
    \vertex [right=0.7cm of a3] (a4);
    \vertex [right=0.7cm of a4] (a5);
    \vertex [right=0.5cm of a5] (a6);
    \vertex [right=1.0cm of a6] (a7);
    \vertex [right=0.5cm of a7] (a8);
    \vertex [right=0.7cm of a8] (a9);
    \vertex [right=0.7cm of a9] (a10);
    \vertex [right=0.7cm of a10] (a11);
    \vertex [right=0.5cm of a11] (a12);    \diagram*{
      (a1) --[plain] (a6),  
      (a2) -- [scalar, out=90, in=90, looseness=1.5,thick] (a4),
      (a3) -- [scalar, out=90, in=90, looseness=1.5,thick] (a5),
      (a7) --[plain] (a12),  
      (a8) -- [scalar, out=90, in=90, looseness=1.5,thick] (a10),
      (a9) -- [scalar, out=90, in=90, looseness=1.5,thick] (a11),
    };
\end{feynman}
\end{tikzpicture}
\end{equation}
\begin{equation}\label{eqn:diagram-4c-xy}
\begin{tikzpicture}
\node at (7.3,-0.3) {\( t,x \)};
\node at (6.7,-0.3) {\( t_1 \)};
\node at (6.1,-0.3) {\( t_2 \)};
\node at (5.4,-0.3) {\( t_3 \)};
\node at (4.8,-0.3) {\( t_4 \)};
\node at (4.2,-0.3) {\( 0,y \)};
\node at (3.6,-0.) {\( =~-\)};
\node at (3.1,-0.3) {\( t,y \)};
\node at (2.6,-0.3) {\( t_1 \)};
\node at (1.9,-0.3) {\( t_2 \)};
\node at (1.2,-0.3) {\( t_3 \)};
\node at (0.6,-0.3) {\( t_4 \)};
\node at (0.,-0.3) {\( 0,x \)};
\node at (7.7,-1.05) {\( =\displaystyle ~\left(\frac{\delta\Omega^2}{3}\right)^2\displaystyle\int_0^t\dif t_1\int_0^{t_1}\dif t_2\int_0^{t_2}\dif t_3\int_0^{t_3}\dif t_4~\varphi(t_1-t_3)\varphi(t_2-t_4) \)};
\node at (7.8,-2.05) {\( \times\displaystyle ~\bigg( -\sin{2\Omega_0(t_1-t_3)}-\sin{2\Omega_0(t_2-t_4)} + \sin{\Omega_0(t_1+t_2-(t_3+t_4))}[1+2\cos{\Omega_0(t_2+t_3-(t_1+t_4))}] \bigg). \)};
\begin{feynman}
    \vertex (a1);
    \vertex [right=0.5cm of a1] (a2);
    \vertex [right=0.7cm of a2] (a3);
    \vertex [right=0.7cm of a3] (a4);
    \vertex [right=0.7cm of a4] (a5);
    \vertex [right=0.5cm of a5] (a6);
    \vertex [right=1.0cm of a6] (a7);
    \vertex [right=0.5cm of a7] (a8);
    \vertex [right=0.7cm of a8] (a9);
    \vertex [right=0.7cm of a9] (a10);
    \vertex [right=0.7cm of a10] (a11);
    \vertex [right=0.5cm of a11] (a12);    \diagram*{
      (a1) --[plain] (a6),  
      (a2) -- [scalar, out=90, in=90, looseness=1.5,thick] (a4),
      (a3) -- [scalar, out=90, in=90, looseness=1.5,thick] (a5),
      (a7) --[plain] (a12),  
      (a8) -- [scalar, out=90, in=90, looseness=1.5,thick] (a10),
      (a9) -- [scalar, out=90, in=90, looseness=1.5,thick] (a11),
    };
\end{feynman}
\end{tikzpicture}
\end{equation}
In the Laplace reciprocal space, these diagrams read in terms of 3D numerical integrations as
\begin{equation}\label{eqn:diagram-4c-xx-laplace}
\begin{tikzpicture}
\node at (7.2,-0.3) {\( y \)};
\node at (6.6,-0.3) {\( t_1 \)};
\node at (6.0,-0.3) {\( t_2 \)};
\node at (5.3,-0.3) {\( t_3 \)};
\node at (4.7,-0.3) {\( t_4 \)};
\node at (4.1,-0.3) {\( y \)};
\node at (3.6,-0.) {\( = \)};
\node at (3.1,-0.3) {\( x \)};
\node at (2.6,-0.3) {\( t_1 \)};
\node at (1.9,-0.3) {\( t_2 \)};
\node at (1.2,-0.3) {\( t_3 \)};
\node at (0.6,-0.3) {\( t_4 \)};
\node at (0.,-0.3) {\( x \)};
\node at (8.4,-1.05) {\( =\displaystyle ~\left(\frac{\delta\Omega^2}{3}\right)^2\mathcal{L}^2[1](s)\displaystyle\iiint\dif x_1\dif x_2\dif x_3~\mathrm{e}^{-s(x_1+x_2+x_3)}\varphi(x_1+x_2)\varphi(x_2+x_3) \)};
\node at (8.8,-2.05) {\( \times\displaystyle ~\left[1+\left( 1-2\cos{\Omega_0(x_1-x_3)}+2\cos{\Omega_0x_3}\right) \cos{\Omega_0(x_1+2x_2+x_3)}  \right]. \)};
\begin{feynman}
    \vertex (a1);
    \vertex [right=0.5cm of a1] (a2);
    \vertex [right=0.7cm of a2] (a3);
    \vertex [right=0.7cm of a3] (a4);
    \vertex [right=0.7cm of a4] (a5);
    \vertex [right=0.5cm of a5] (a6);
    \vertex [right=1.0cm of a6] (a7);
    \vertex [right=0.5cm of a7] (a8);
    \vertex [right=0.7cm of a8] (a9);
    \vertex [right=0.7cm of a9] (a10);
    \vertex [right=0.7cm of a10] (a11);
    \vertex [right=0.5cm of a11] (a12);    \diagram*{
      (a1) --[plain] (a6),  
      (a2) -- [scalar, out=90, in=90, looseness=1.5,thick] (a4),
      (a3) -- [scalar, out=90, in=90, looseness=1.5,thick] (a5),
      (a7) --[plain] (a12),  
      (a8) -- [scalar, out=90, in=90, looseness=1.5,thick] (a10),
      (a9) -- [scalar, out=90, in=90, looseness=1.5,thick] (a11),
    };
\end{feynman}
\end{tikzpicture}
\end{equation}
\begin{equation}\label{eqn:diagram-4c-xy}
\begin{tikzpicture}
\node at (7.3,-0.3) {\( x \)};
\node at (6.7,-0.3) {\( t_1 \)};
\node at (6.1,-0.3) {\( t_2 \)};
\node at (5.4,-0.3) {\( t_3 \)};
\node at (4.8,-0.3) {\( t_4 \)};
\node at (4.2,-0.3) {\( y \)};
\node at (3.6,-0.) {\( =~-\)};
\node at (3.1,-0.3) {\( y \)};
\node at (2.6,-0.3) {\( t_1 \)};
\node at (1.9,-0.3) {\( t_2 \)};
\node at (1.2,-0.3) {\( t_3 \)};
\node at (0.6,-0.3) {\( t_4 \)};
\node at (0.,-0.3) {\( x \)};
\node at (8.2,-1.05) {\( =\displaystyle ~\left(\frac{\delta\Omega^2}{3}\right)^2\mathcal{L}^2[1](s)\displaystyle\iiint\dif x_1\dif x_2\dif x_3~\mathrm{e}^{-s(x_1+x_2+x_3)}\varphi(x_1+x_2)\varphi(x_2+x_3) \)};
\node at (8.8,-2.05) {\( \times\displaystyle ~\left[\left( 1-2\cos{\Omega_0(x_1-x_3)}+2\cos{\Omega_0x_3}\right) \sin{\Omega_0(x_1+2x_2+x_3)}  \right]. \)};
\begin{feynman}
    \vertex (a1);
    \vertex [right=0.5cm of a1] (a2);
    \vertex [right=0.7cm of a2] (a3);
    \vertex [right=0.7cm of a3] (a4);
    \vertex [right=0.7cm of a4] (a5);
    \vertex [right=0.5cm of a5] (a6);
    \vertex [right=1.0cm of a6] (a7);
    \vertex [right=0.5cm of a7] (a8);
    \vertex [right=0.7cm of a8] (a9);
    \vertex [right=0.7cm of a9] (a10);
    \vertex [right=0.7cm of a10] (a11);
    \vertex [right=0.5cm of a11] (a12);    \diagram*{
      (a1) --[plain] (a6),  
      (a2) -- [scalar, out=90, in=90, looseness=1.5,thick] (a4),
      (a3) -- [scalar, out=90, in=90, looseness=1.5,thick] (a5),
      (a7) --[plain] (a12),  
      (a8) -- [scalar, out=90, in=90, looseness=1.5,thick] (a10),
      (a9) -- [scalar, out=90, in=90, looseness=1.5,thick] (a11),
    };
\end{feynman}
\end{tikzpicture}
\end{equation}

The calculation of the nested diagrams proceeds the same manner. The general expression reads as
\begin{equation}\label{eqn:diagram-4c-xx}
\begin{tikzpicture}
\node at (3.1,-0.3) {\( t,z \)};
\node at (2.6,-0.3) {\( t_1 \)};
\node at (1.9,-0.3) {\( t_2 \)};
\node at (1.2,-0.3) {\( t_3 \)};
\node at (0.6,-0.3) {\( t_4 \)};
\node at (0.,-0.3) {\( 0,z \)};
\node at (7.8,0.05) {\( =\displaystyle ~\left(\frac{\delta\Omega^2}{3}\right)^2\displaystyle\int_0^t\dif t_1\int_0^{t_1}\dif t_2\int_0^{t_2}\dif t_3\int_0^{t_3}\dif t_4~\varphi(t_1-t_4)\varphi(t_2-t_3) \)};
\node at (7.8,-0.95) {\( \times\displaystyle ~\left( R_{i_0k_1}(\Omega_0t_1) R_{i_3k_1}(\Omega_0t_4)R^{-1}_{m_1i_1}(\Omega_0t_1)R^{-1}_{m_1i_4}(\Omega_0t_4)- R_{i_0k_1}(\Omega_0t_1) R_{i_3k_4}(\Omega_0t_4)R^{-1}_{k_4i_1}(\Omega_0t_1)R^{-1}_{k_1i_4}(\Omega_0t_4)\right) \)};
\node at (7.9,-1.95) {\( \times\displaystyle ~\left( R_{i_1k_2}(\Omega_0t_2) R_{i_2k_2}(\Omega_0t_3)R^{-1}_{m_2i_2}(\Omega_0t_2)R^{-1}_{m_2i_3}(\Omega_0t_3)- R_{i_1k_2}(\Omega_0t_2) R_{i_2k_3}(\Omega_0t_3)R^{-1}_{k_3i_2}(\Omega_0t_2)R^{-1}_{k_2i_3}(\Omega_0t_3)\right), \)};
\begin{feynman}
    \vertex (a1);
    \vertex [right=0.5cm of a1] (a2);
    \vertex [right=0.7cm of a2] (a3);
    \vertex [right=0.7cm of a3] (a4);
    \vertex [right=0.7cm of a4] (a5);
    \vertex [right=0.5cm of a5] (a6);
    \diagram*{
      (a1) --[plain] (a6),  
      (a2) -- [scalar, out=90, in=90, looseness=1.5,thick] (a5),
      (a3) -- [scalar, out=90, in=90, looseness=2.0,thick] (a4),
    };
\end{feynman}
\end{tikzpicture}
\end{equation}
which leads to 
\begin{equation}\label{eqn:diagram-4n-xx}
\begin{tikzpicture}
\node at (7.1,-0.3) {\( t,y \)};
\node at (6.5,-0.3) {\( t_1 \)};
\node at (5.9,-0.3) {\( t_2 \)};
\node at (5.3,-0.3) {\( t_3 \)};
\node at (4.7,-0.3) {\( t_4 \)};
\node at (4.1,-0.3) {\( 0,y \)};
\node at (3.6,-0.) {\( = \)};
\node at (3.1,-0.3) {\( t,x \)};
\node at (2.6,-0.3) {\( t_1 \)};
\node at (1.9,-0.3) {\( t_2 \)};
\node at (1.2,-0.3) {\( t_3 \)};
\node at (0.6,-0.3) {\( t_4 \)};
\node at (0.,-0.3) {\( 0,x \)};
\node at (7.85,-1.05) {\( =\displaystyle ~\left(\frac{\delta\Omega^2}{3}\right)^2\displaystyle\int_0^t\dif t_1\int_0^{t_1}\dif t_2\int_0^{t_2}\dif t_3\int_0^{t_3}\dif t_4~\varphi(t_1-t_4)\varphi(t_2-t_3) \)};
\node at (7.8,-2.05) {\( \times\displaystyle ~\bigg( \cos{2\Omega_0(t_1+t_3-(t_2+t_4))}-\cos{\Omega_0(2(t_1-t_4)+t_3-t_2)}[1+2\cos{\Omega_0(t_2-t_3)}] \)};
\node at (7.8,-3.05) {\( -\cos{\Omega_0(t_1-t_4)}[1+2\cos{\Omega_0(2(t_3-t_2)+t_1-t_4)}] + \cos{\Omega_0(t_1-t_4)}[1+2\cos{\Omega_0(t_1+t_3-(t_2+t_4))}][1+2\cos{\Omega_0(t_2-t_3)}] \bigg), \)};
\begin{feynman}
    \vertex (a1);
    \vertex [right=0.5cm of a1] (a2);
    \vertex [right=0.7cm of a2] (a3);
    \vertex [right=0.7cm of a3] (a4);
    \vertex [right=0.7cm of a4] (a5);
    \vertex [right=0.5cm of a5] (a6);
    \vertex [right=1.0cm of a6] (a7);
    \vertex [right=0.5cm of a7] (a8);
    \vertex [right=0.7cm of a8] (a9);
    \vertex [right=0.7cm of a9] (a10);
    \vertex [right=0.7cm of a10] (a11);
    \vertex [right=0.5cm of a11] (a12);    \diagram*{
      (a1) --[plain] (a6),  
      (a2) -- [scalar, out=90, in=90, looseness=1.5,thick] (a5),
      (a3) -- [scalar, out=90, in=90, looseness=2.0,thick] (a4),
      (a7) --[plain] (a12),  
      (a8) -- [scalar, out=90, in=90, looseness=1.5,thick] (a11),
      (a9) -- [scalar, out=90, in=90, looseness=2.0,thick] (a10),
    };
\end{feynman}
\end{tikzpicture}
\end{equation}
and to
\begin{equation}\label{eqn:diagram-4n-xy}
\begin{tikzpicture}
\node at (7.2,-0.3) {\( t,x \)};
\node at (6.6,-0.3) {\( t_1 \)};
\node at (6.0,-0.3) {\( t_2 \)};
\node at (5.4,-0.3) {\( t_3 \)};
\node at (4.8,-0.3) {\( t_4 \)};
\node at (4.2,-0.3) {\( 0,y \)};
\node at (3.6,-0.) {\( =~- \)};
\node at (3.1,-0.3) {\( t,y \)};
\node at (2.6,-0.3) {\( t_1 \)};
\node at (1.9,-0.3) {\( t_2 \)};
\node at (1.2,-0.3) {\( t_3 \)};
\node at (0.6,-0.3) {\( t_4 \)};
\node at (0.,-0.3) {\( 0,x \)};
\node at (7.7,-1.05) {\( =\displaystyle ~\left(\frac{\delta\Omega^2}{3}\right)^2\displaystyle\int_0^t\dif t_1\int_0^{t_1}\dif t_2\int_0^{t_2}\dif t_3\int_0^{t_3}\dif t_4~\varphi(t_1-t_4)\varphi(t_2-t_3) \)};
\node at (7.8,-2.05) {\( \times\displaystyle ~\bigg( \sin{2\Omega_0(t_1+t_3-(t_2+t_4))}-\sin{\Omega_0(2(t_1-t_4)+t_3-t_2)}[1+2\cos{\Omega_0(t_2-t_3)}] \)};
\node at (7.8,-3.05) {\( -\sin{\Omega_0(t_1-t_4)}[1+2\cos{\Omega_0(2(t_3-t_2)+t_1-t_4)}] + \sin{\Omega_0(t_1-t_4)}[1+2\cos{\Omega_0(t_1+t_3-(t_2+t_4))}][1+2\cos{\Omega_0(t_2-t_3)}] \bigg). \)};
\begin{feynman}
    \vertex (a1);
    \vertex [right=0.5cm of a1] (a2);
    \vertex [right=0.7cm of a2] (a3);
    \vertex [right=0.7cm of a3] (a4);
    \vertex [right=0.7cm of a4] (a5);
    \vertex [right=0.5cm of a5] (a6);
    \vertex [right=1.0cm of a6] (a7);
    \vertex [right=0.5cm of a7] (a8);
    \vertex [right=0.7cm of a8] (a9);
    \vertex [right=0.7cm of a9] (a10);
    \vertex [right=0.7cm of a10] (a11);
    \vertex [right=0.5cm of a11] (a12);    \diagram*{
      (a1) --[plain] (a6),  
      (a2) -- [scalar, out=90, in=90, looseness=1.5,thick] (a5),
      (a3) -- [scalar, out=90, in=90, looseness=2.0,thick] (a4),
      (a7) --[plain] (a12),  
      (a8) -- [scalar, out=90, in=90, looseness=1.5,thick] (a11),
      (a9) -- [scalar, out=90, in=90, looseness=2.0,thick] (a10),
    };
\end{feynman}
\end{tikzpicture}
\end{equation}
In the Laplace space, the diagrams read as
\begin{equation}\label{eqn:diagram-4n-xx-laplace}
\begin{tikzpicture}
\node at (7.1,-0.3) {\( y \)};
\node at (6.5,-0.3) {\( t_1 \)};
\node at (5.9,-0.3) {\( t_2 \)};
\node at (5.3,-0.3) {\( t_3 \)};
\node at (4.7,-0.3) {\( t_4 \)};
\node at (4.1,-0.3) {\( y \)};
\node at (3.6,-0.) {\( = \)};
\node at (3.1,-0.3) {\( x \)};
\node at (2.6,-0.3) {\( t_1 \)};
\node at (1.9,-0.3) {\( t_2 \)};
\node at (1.2,-0.3) {\( t_3 \)};
\node at (0.6,-0.3) {\( t_4 \)};
\node at (0.,-0.3) {\( x \)};
\node at (8.4,-1.05) {\( =\displaystyle ~\left(\frac{\delta\Omega^2}{3}\right)^2\mathcal{L}^2[1](s)\displaystyle\iiint\dif x_1\dif x_2\dif x_3~\mathrm{e}^{-s(x_1+x_2+x_3)}\varphi(x_1+x_2+x_3)\varphi(x_2) \)};
\node at (8.8,-2.05) {\( \times\displaystyle ~\left( 1+\cos{\Omega_0x_2}+\cos{\Omega_0(x_1+x_3)}+\cos{\Omega_0(x_1+2x_2+x_3)} \right) , \)};
\begin{feynman}
    \vertex (a1);
    \vertex [right=0.5cm of a1] (a2);
    \vertex [right=0.7cm of a2] (a3);
    \vertex [right=0.7cm of a3] (a4);
    \vertex [right=0.7cm of a4] (a5);
    \vertex [right=0.5cm of a5] (a6);
    \vertex [right=1.0cm of a6] (a7);
    \vertex [right=0.5cm of a7] (a8);
    \vertex [right=0.7cm of a8] (a9);
    \vertex [right=0.7cm of a9] (a10);
    \vertex [right=0.7cm of a10] (a11);
    \vertex [right=0.5cm of a11] (a12);    \diagram*{
      (a1) --[plain] (a6),  
      (a2) -- [scalar, out=90, in=90, looseness=1.5,thick] (a5),
      (a3) -- [scalar, out=90, in=90, looseness=2.0,thick] (a4),
      (a7) --[plain] (a12),  
      (a8) -- [scalar, out=90, in=90, looseness=1.5,thick] (a11),
      (a9) -- [scalar, out=90, in=90, looseness=2.0,thick] (a10),
    };
\end{feynman}
\end{tikzpicture}
\end{equation}
\begin{equation}\label{eqn:diagram-4n-xy-laplace}
\begin{tikzpicture}
\node at (7.2,-0.3) {\( x \)};
\node at (6.6,-0.3) {\( t_1 \)};
\node at (6.0,-0.3) {\( t_2 \)};
\node at (5.4,-0.3) {\( t_3 \)};
\node at (4.8,-0.3) {\( t_4 \)};
\node at (4.2,-0.3) {\( y \)};
\node at (3.6,-0.) {\( =~- \)};
\node at (3.1,-0.3) {\( y \)};
\node at (2.6,-0.3) {\( t_1 \)};
\node at (1.9,-0.3) {\( t_2 \)};
\node at (1.2,-0.3) {\( t_3 \)};
\node at (0.6,-0.3) {\( t_4 \)};
\node at (0.,-0.3) {\( x \)};
\node at (8.2,-1.05) {\( =\displaystyle ~\left(\frac{\delta\Omega^2}{3}\right)^2\mathcal{L}^2[1](s)\displaystyle\iiint\dif x_1\dif x_2\dif x_3~\mathrm{e}^{-s(x_1+x_2+x_3)}\varphi(x_1+x_2+x_3)\varphi(x_2) \)};
\node at (7.8,-2.05) {\( \times\displaystyle ~\left( \sin{\Omega_0x_2}+2\cos{\Omega_0x_2}\sin{\Omega_0(x_1+x_2+x_3}) \right).\)};
\begin{feynman}
    \vertex (a1);
    \vertex [right=0.5cm of a1] (a2);
    \vertex [right=0.7cm of a2] (a3);
    \vertex [right=0.7cm of a3] (a4);
    \vertex [right=0.7cm of a4] (a5);
    \vertex [right=0.5cm of a5] (a6);
    \vertex [right=1.0cm of a6] (a7);
    \vertex [right=0.5cm of a7] (a8);
    \vertex [right=0.7cm of a8] (a9);
    \vertex [right=0.7cm of a9] (a10);
    \vertex [right=0.7cm of a10] (a11);
    \vertex [right=0.5cm of a11] (a12);    \diagram*{
      (a1) --[plain] (a6),  
      (a2) -- [scalar, out=90, in=90, looseness=1.5,thick] (a5),
      (a3) -- [scalar, out=90, in=90, looseness=2.0,thick] (a4),
      (a7) --[plain] (a12),  
      (a8) -- [scalar, out=90, in=90, looseness=1.5,thick] (a11),
      (a9) -- [scalar, out=90, in=90, looseness=2.0,thick] (a10),
    };
\end{feynman}
\end{tikzpicture}
\end{equation}

\end{widetext}

\section{Red-noise approximation}
\label{app:red-noise}

\begin{figure}[h]
\centering
\includegraphics[width=\columnwidth]{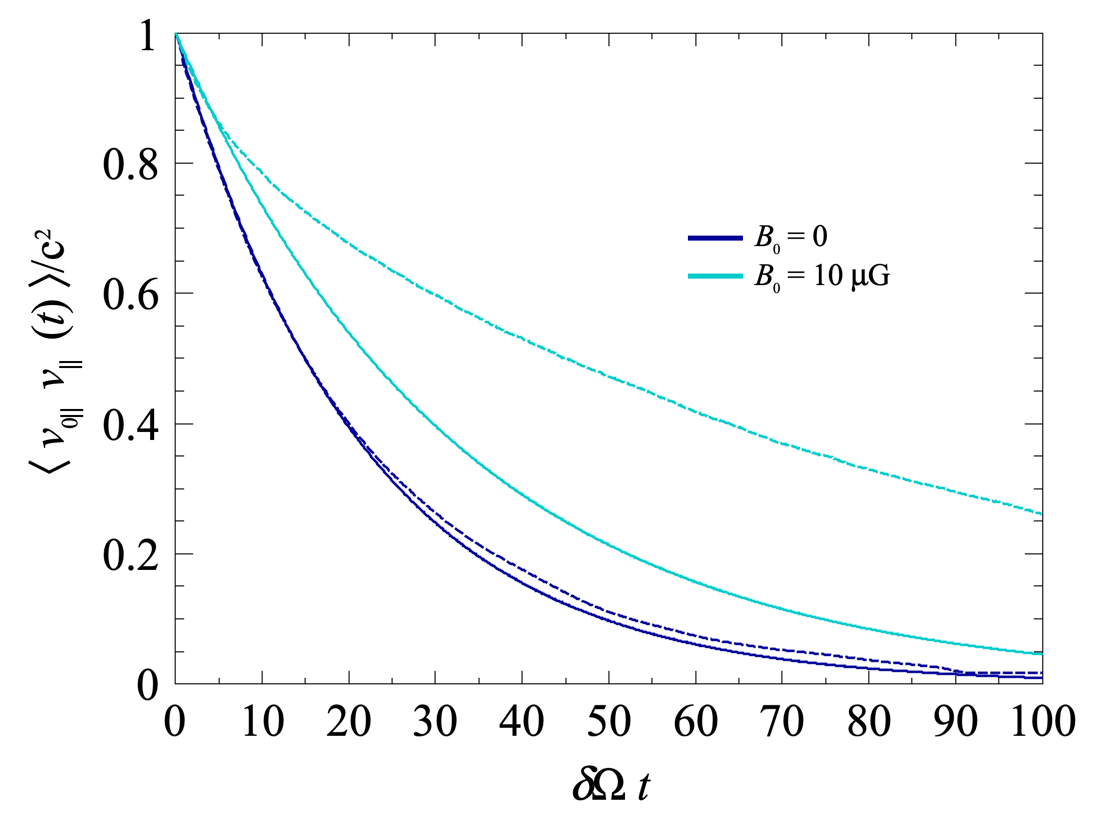}
\caption{Parallel velocity decorrelation function of particles with reduced rigidity $\rho=1$ for $\delta B=1~\upmu$G and different values of $B_0$ as a function of the gyroperiod time scale $\delta\Omega t$, as obtained from the red-noise approximation. Dashed lines are from Monte-Carlo simulations.}
\label{fig:rn}
\end{figure}

\begin{figure}[h]
\centering
\includegraphics[width=\columnwidth]{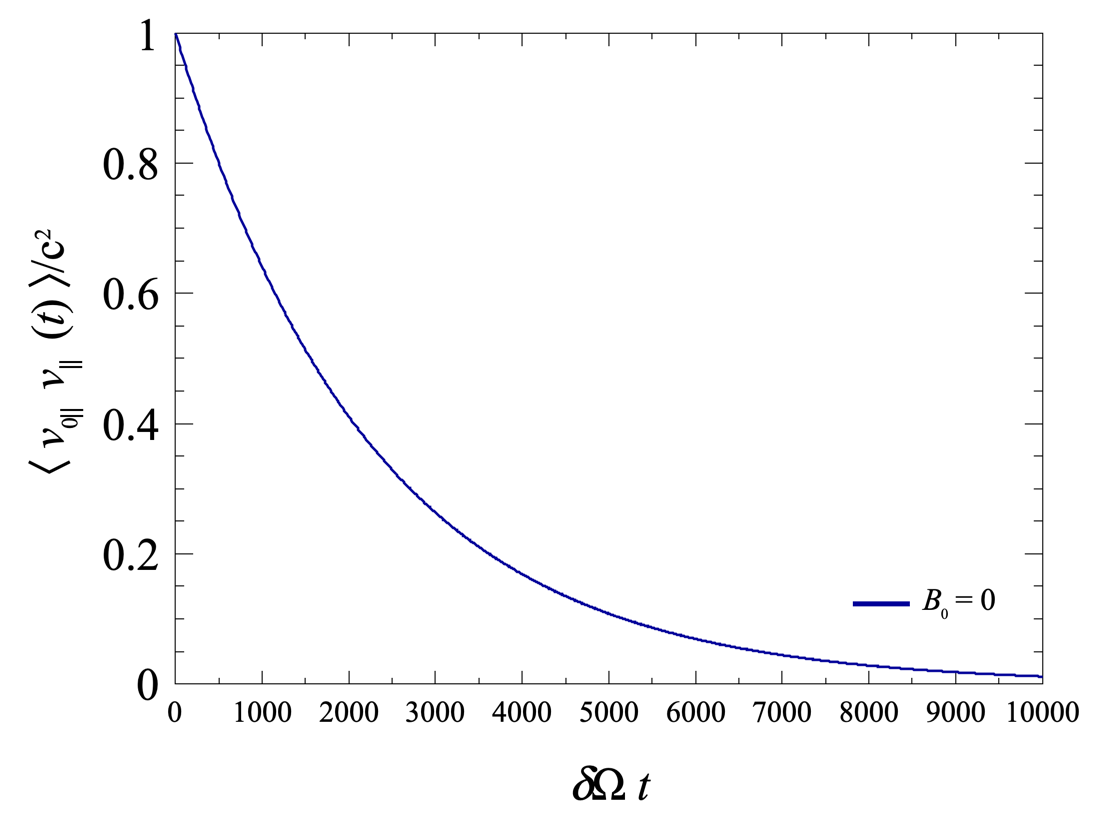}
\caption{Same as Fig.~\ref{fig:rn} for $\rho=10$, $\delta B=1~\upmu$G and $B_0=0$.}
\label{fig:rn2}
\end{figure}

We present in this Appendix results in the context of modeling $\varphi(t)$ as a red-noise process with parameter $\tau$, 
\begin{equation}
\label{eqn:phi}
\varphi(t)=\exp{(-t/\tau)}.
\end{equation}
Thanks to the property of the exponential function $\varphi(t_1+t_2)=\varphi(t_1)\varphi(t_2)$, analytical results (in the Laplace reciprocal space) can be obtained for $\hat{W}_{0i}(s)$. We restrict the range of application of the results presented in this Appendix to the quasi-ballistic regime (that is, $\rho\geq0.5$) for which an heuristic expression for $\tau$ is
\begin{equation}
\label{eqn:tau}
\tau(\rho)\simeq \frac{L_\mathrm{max}}{16~\rho c}.
\end{equation}
Treating the case of the parallel diffusion, the contribution of the unconnected diagrams to the zigzag propagator reads as
\begin{equation}\label{eqn:bourret-2-laplace-rn}
\begin{tikzpicture}
\node at (2.4,-0.3) {\( z \)};
\node at (1.7,-0.3) {\( t_1 \)};
\node at (0.8,-0.3) {\( t_2 \)};
\node at (0.,-0.3) {\( z \)};
\node at (5.2,0.05) {\( = \displaystyle ~-2\frac{\delta\Omega^2}{3}\frac{\hat{W}_{0z}(s)}{s}
\frac{\tau(1+s\tau)}{(1+s\tau)^2+(\Omega_0\tau)^2}. \)};
\begin{feynman}
    \vertex (a1);
    \vertex [right=0.7cm of a1] (a2);
    \vertex [right=1cm of a2] (a3);
    \vertex [right=0.7cm of a3] (a4);
    \diagram*{
      (a1) --[plain] (a2),  
      (a2) --[plain] (a3),  
      (a3) --[boson] (a4),  
      (a2) -- [scalar, out=90, in=90, looseness=2.0,thick] (a3),
    };
  \end{feynman}
\end{tikzpicture}
\end{equation}
To explore the impact that an improved zigzag propagator might have, we also consider additional classes of diagrams to approximate the mass propagator that would be obtained by substituting the thin line between $t_2$ and $t_1$ for a zigzag (summation of all nested diagrams). To do so, we add to Eqn.~\ref{eqn:bourret} the following contributions:
\begin{widetext}
\begin{equation}\label{eqn:2loops}
\begin{tikzpicture}
\node at (3.0,-0.3) {\( z \)};
\node at (2.4,-0.3) {\( t_1 \)};
\node at (1.8,-0.3) {\( t_2 \)};
\node at (1.2,-0.3) {\( t_3 \)};
\node at (0.6,-0.3) {\( t_4 \)};
\node at (0,-0.3) {\( z \)};
\node at (10.5,0.) {\( =~\displaystyle 2\tau^3\frac{\delta\Omega^2}{3}\frac{\hat{W}_{0z}(s)}{s} \frac{8+12(s\tau)^3+2(s\tau)^4-11(\Omega_0\tau)^2-(\Omega_0\tau)^4-12s\tau(-2+(\Omega_0\tau)^2)+s^2(26\tau^2-3\Omega_0^2\tau^4)}{(2+s\tau)((1+s\tau)^2+(\Omega_0\tau)^2)^2((2+s\tau)^2+(\Omega_0\tau)^2)},\)};
\begin{feynman}
    \vertex (a1);
    \vertex [right=0.6cm of a1] (a2);
    \vertex [right=0.6cm of a2] (a3);
    \vertex [right=0.6cm of a3] (a4);
    \vertex [right=0.6cm of a4] (a5);
    \vertex [right=0.6cm of a5] (a6);
    \diagram*{
      (a1) --[plain] (a5),      
      (a5) --[boson] (a6),    
      (a2) -- [scalar, out=90, in=90, looseness=1.5,thick] (a5),
      (a3) -- [scalar, out=90, in=90, looseness=2.0,thick] (a4),
    };
  \end{feynman}
\end{tikzpicture}
\end{equation}

\begin{equation}\label{eqn:3loops}
\begin{tikzpicture}
\node at (4.2,-0.3) {\( z \)};
\node at (3.6,-0.3) {\( t_1 \)};
\node at (3.0,-0.3) {\( t_2 \)};
\node at (2.4,-0.3) {\( t_3 \)};
\node at (1.8,-0.3) {\( t_4 \)};
\node at (1.2,-0.3) {\( t_5 \)};
\node at (0.6,-0.3) {\( t_6 \)};
\node at (0,-0.3) {\( z \)};
\node at (10.8,0.) {\( =~ \displaystyle 2\tau^5\frac{\delta\Omega^2}{3}\frac{\hat{W}_{0z}(s)}{s} \frac{1}{(3+s\tau)((1+s\tau)^2+(\Omega_0\tau)^2)^2((2+s\tau)^2+(\Omega_0\tau)^2)^2((3+s\tau)^2+(\Omega_0\tau)^2)} \)};
\node at (10.8,-1.) {\( \times\displaystyle \bigg[2(1+s\tau)^2(2+s\tau)^2(3+s\tau)^2+(\Omega_0\tau) ^2(70+178s\tau+139(s\tau)^2  \)};
\node at (11.8,-2.) {\( \displaystyle +44(s\tau)^3+5(s\tau)^4+27(\Omega_0\tau)^2+20s\tau^3\Omega_0^2+4s^2\tau^4\Omega_0^2+(\Omega_0\tau)^4)\bigg]. \)};
\begin{feynman}
    \vertex (a1);
    \vertex [right=0.6cm of a1] (a2);
    \vertex [right=0.6cm of a2] (a3);
    \vertex [right=0.6cm of a3] (a4);
    \vertex [right=0.6cm of a4] (a5);
    \vertex [right=0.6cm of a5] (a6);
    \vertex [right=0.6cm of a6] (a7);
    \vertex [right=0.6cm of a7] (a8);
    \diagram*{
      (a1) --[plain] (a7),      
      (a7) --[boson] (a8),    
      (a2) -- [scalar, out=90, in=90, looseness=1.1,thick] (a7),
      (a3) -- [scalar, out=90, in=90, looseness=1.2,thick] (a6),
      (a4) -- [scalar, out=90, in=90, looseness=2.0,thick] (a5),
    };
  \end{feynman}
\end{tikzpicture}
\end{equation}
The expressions for the iterated propagator require only 1D integrations:
\begin{equation}\label{eqn:iter-n-laplace-rn}
\begin{tikzpicture}
\node at (2.4,-0.3) {\( z \)};
\node at (1.7,-0.3) {\( t_1 \)};
\node at (0.8,-0.3) {\( t_2 \)};
\node at (0.,-0.3) {\( z \)};
\node at (6.0,0.05) {\( =~\displaystyle -2\frac{\delta\Omega^2}{3}\frac{\hat{W}_{0z}(s)}{s}\mathcal{L}[\mathrm{e}^{-x/\tau}\langle w_{0z}(x)\rangle_0\cos{\Omega_0x}](s), \)};
\begin{feynman}
    \vertex (a1);
    \vertex [right=0.7cm of a1] (a2);
    \vertex [right=1cm of a2] (a3);
    \vertex [right=0.7cm of a3] (a4);
    \diagram*{
      (a1) --[plain] (a2),  
      (a2) --[boson] (a3),  
      (a3) --[double] (a4),  
      (a2) -- [scalar, out=90, in=90, looseness=2.0,thick] (a3),
    };
  \end{feynman}
\end{tikzpicture}
\end{equation}
\begin{equation}\label{eqn:iter-c-laplace-rn}
\begin{tikzpicture}
\node at (3.6,-0.3) {\( z \)};
\node at (2.9,-0.3) {\( t_1 \)};
\node at (2.2,-0.3) {\( t_2 \)};
\node at (1.5,-0.3) {\( t_3 \)};
\node at (0.8,-0.3) {\( t_4 \)};
\node at (0.,-0.3) {\( z \)};
\node at (7.0,0){\( =~\displaystyle 2\left(\frac{\delta\Omega^2}{3}\right)^2\frac{\hat{W}_{0z}(s)}{s}\mathcal{L}[\mathrm{e}^{-2x/\tau}\langle w_{0z}(x)\rangle_0](s) \)};
\node at (9.7,-1){\( ~\displaystyle \times \left(\mathcal{L}^2[\mathrm{e}^{-x/\tau}\langle w_{0z}(x)\rangle_0\cos{\Omega_0x}](s) +\mathcal{L}^2[\mathrm{e}^{-x/\tau}\langle w_{0z}(x)\rangle_0\sin{\Omega_0x}](s) \right). \)};
\begin{feynman}
    \vertex (a1);
    \vertex [right=0.7cm of a1] (a2);
    \vertex [right=0.7cm of a2] (a3);
    \vertex [right=0.7cm of a3] (a4);  \vertex [right=0.7cm of a4] (a5);
    \vertex [right=0.7cm of a5] (a6);{~=~};
    \diagram*{
      (a1) --[plain] (a2),  
      (a2) --[boson] (a5),  
      (a5) --[double] (a6),  
      (a2) -- [scalar, out=90, in=90, looseness=1.5,thick] (a4),
      (a3) -- [scalar, out=90, in=90, looseness=1.5,thick] (a5),
    };
  \end{feynman}
\end{tikzpicture}
\end{equation}
\end{widetext}
Results are displayed in Fig.~\ref{fig:rn} for $\rho=1$ and different values of $B_0$ (the case $B_0=100~\upmu$G is not shown as the calculation yields to nonphysical results with regions such that $\langle v_{0\parallel}v_\parallel(t)\rangle/c^2>1$). Also shown in Fig.~\ref{fig:rn2} is the decorrelation function obtained for $\rho=10$: the (exponential) decay is observed to be $\sim 100$ times that obtained for $\rho=1$, yielding a diffusion coefficient scaling effectively as $\rho^2$, as expected. We note that including or not the classes of diagrams depicted in Eqn.~\ref{eqn:2loops} and Eqn.~\ref{eqn:3loops} changes the results very little; this reinforces the soundness of Eqn.~\ref{eqn:bourret} to approximate the mass propagator during the zeroth iteration. Overall, the red-noise approximation for the 2-pt function of the turbulence experienced by particles between two successive times is therefore competitive in the high-rigidity regime; yet the range applicability for $B_0$ is reduced.

\bibliographystyle{apsrev4-2}
\bibliography{bibliography.bib}

\end{document}